\documentclass[dvipdfmx]{ptephy}

\usepackage{graphics} 

\usepackage[normalem]{ulem}  
\usepackage{natbib} 

\usepackage{amsmath} 
\usepackage{amsthm} 

\usepackage{bm}

\renewcommand{\sout}{\bgroup \color[rgb]{1,0,0} \ULdepth=-.5ex \ULset}

\newcommand{\Psfig}[2]{\includegraphics[width=#1]{#2}}
\newcommand{\PsfigII}[2]{\includegraphics[scale=#1]{#2}}

\def\mev{\text{ MeV}}
\def\gev{\text{ GeV}}
\def\fm{\text{ fm}}
\def\microb{~\mu \text{b}}

\def\Kaellen{K\"{a}llen }

\def\HeT{{}^{3} \text{He}}

\begin{document}

\title{On the structure observed in the in-flight ${}^{3}\text{He} (
  K^{-} , \, \Lambda p ) n$ reaction at J-PARC}

\author{\name{Takayasu~Sekihara}{1,\ast}
  \name{Eulogio~Oset}{2},
  and 
  \name{Angels~Ramos}{3}
}

\address{ \affil{1}{Advanced Science Research Center, Japan Atomic
    Energy Agency, Shirakata, Tokai, Ibaraki, 319-1195, Japan} %
  \affil{2}{Departamento de F\'{\i}sica Te\'orica and IFIC, Centro
    Mixto Universidad de Valencia-CSIC, Institutos de Investigaci\'on
    de Paterna, Aptdo. 22085, 46071 Valencia, Spain} %
  \affil{3}{Departament de F\'{\i}sica Qu\`antica i Astrof\'{\i}sica
    and Institut de Ci\`encies del Cosmos, Universitat de Barcelona,
    Mart\'i i Franqu\`es 1, 08028 Barcelona, Spain} %
  \email{sekihara@post.j-parc.jp}
}

\begin{abstract}%

  A theoretical investigation is done to clarify the origin of the
  peak structure observed near the $K^{-} p p$ threshold in the
  in-flight $\HeT (K^{-}, \, \Lambda p) n$ reaction of the J-PARC E15
  experiment, which could be a signal of the lightest kaonic nuclei,
  that is, the $\bar{K} N N (I=1/2)$ state.  For the investigation, we
  evaluate the $\Lambda p$ invariant mass spectrum assuming two
  possible scenarios to interpret the experimental peak. One assumes
  that the $\Lambda (1405)$ resonance is generated after the emission
  of an energetic neutron from the absorption of the initial $K^-$,
  not forming a bound state with the remaining proton. This
  uncorrelated $\Lambda (1405) p$ system subsequently decays into the
  final $\Lambda p$.  The other scenario implies that, after the
  emission of the energetic neutron, a $\bar{K} N N$ bound state is
  formed, decaying eventually into a $\Lambda p$ pair. Our results
  show that the experimental signal observed in the in-flight $\HeT (
  K^{-} , \, \Lambda p ) n$ reaction at J-PARC is qualitatively well
  reproduced by the assumption that a $\bar{K} N N$ bound state is
  generated in the reaction, definitely discarding the interpretation
  in terms of an uncorrelated $\Lambda (1405) p$ state.
  
\end{abstract}



\maketitle

\section{Introduction}

The study of the $\bar{K} N$ interaction with coupled channels has
been a traditional test field for chiral dynamics in its unitarized
version.  Since the pioneer works of~\cite{Kaiser:1995eg,
  Oset:1997it}, many works have been done in this field including also
the contribution of the higher order chiral Lagrangians (see recent
review in~\cite{Kamiya:2016jqc}).  One of the unexpected results was
the finding of two poles for the $\Lambda (1405)$ resonance
\cite{Oller:2000fj, Jido:2003cb}, which recently became official in
the Particle Data Group~\cite{Agashe:2014kda} [see note on the
$\Lambda (1405)$~\cite{Meissner:2015}].  The other issue that became
topical was the possibility of having kaonic nuclei, in particular a
bound $\bar{K} N N$ system.  The study of this system has been
thoroughly addressed theoretically~\cite{Akaishi:2002bg,
  Shevchenko:2006xy, Shevchenko:2007ke, Ikeda:2007nz, Ikeda:2008ub,
  Ikeda:2010tk, Dote:2008in, Dote:2008hw, Wycech:2008wf,
  Bayar:2011qj,Barnea:2012qa,Dote:2014via} (see review paper
in~\cite{Gal:2013vx}), obtaining a binding energy that varies from a
few MeV to 100 MeV.  There are also discrepancies in the width of the
state, which varies from 10 to 90 MeV.  One step forward in the
evaluation of the width was given in~\cite{Bayar:2012hn}, where
two-nucleon $\bar{K}$ absorption was explicitly considered.  In that
study a binding of 15--30 MeV was found, together with a width of the
order of 80 MeV.  The fact that the width is larger than the binding
energy is shared by most of the theoretical approaches.  One
interesting point of view was to consider this system as a bound state
of $\Lambda (1405) N$ \cite{Uchino:2011jt}.  This allows one to make
some qualitative pictures that help understanding some of the results
obtained when studying the possible formation of this system in
different reactions.  There have been previous claims of the formation
of this state in some experiments, but soon the experimental signals
were interpreted in terms of unavoidable conventional mechanisms (see
overview in \cite{Gal:2013vx, Ramos:2008zza}).  Yet, the experimental
search has continued~\cite{Agnello:2005qj, Yamazaki:2010mu,
  Tokiyasu:2013mwa, Ichikawa:2014ydh, Hashimoto:2014cri} with often
contradictory views (see overview in~\cite{Sada:2016vkt}).

In this line, very recently, a peak structure was observed near the
$K^{-} p p$ threshold in the $\Lambda p$ invariant mass spectrum of
the in-flight $\HeT (K^{-} , \Lambda p) n$ reaction of the J-PARC E15
experiment~\cite{Sada:2016vkt}.  According to their analysis, this
peak can be described by the Breit-Wigner formula with mass $M_{X} =
2355 \, ^{+6}_{-4} \text{(stat.)} \pm 12 \text{(sys.)} \mev$ and width
$\Gamma _{X} = 110 \, ^{+19}_{-17} \text{(stat.)} \pm 27 \text{(sys.)}
\mev$.  This structure could be a signal of the $\bar{K} N N (I=1/2)$
bound state with a binding of $\sim 15 \mev$ from the $K^{-} p p$
threshold.

In this paper we theoretically investigate the origin of the peak
structure observed in the J-PARC E15 experiment.  For this purpose, we
take into account two possible mechanisms for producing a peak in the
mass spectrum of the ${}^{3} \text{He} ( K^{-}, \, \Lambda p ) n$
reaction.  One corresponds to assuming the formation of a $\Lambda
(1405)$ resonance that does not form a bound state with the
remaining proton, while the other considers the formation of a bound
state of the $\bar{K} N N$ system.  We evaluate the cross section of
the ${}^{3} \text{He} ( K^{-}, \, \Lambda p ) n$ reaction assuming the
$\bar{K} N \to \bar{K} N$ scattering around threshold and the
$\Lambda (1405)$ resonance to be described by the chiral unitary
approach~\cite{Kaiser:1995eg, Oset:1997it, Oller:2000fj, Jido:2003cb},
while the description of the $\bar{K} N N$ bound state is done in
terms of the so-called fixed center approximation to the Faddeev
equation~\cite{Bayar:2011qj, Bayar:2012hn}.

As a result, we can unambiguously interpret the experimental spectrum
in the scenario of a $\bar{K} N N$ broad bound state, obtained from
the interaction of the $\bar{K}$ with a pair of
nucleons~\cite{Bayar:2012hn}.  In addition, in this scenario, we
obtain a two peak structure of the mass spectrum near the $\bar{K} N
N$ threshold.  The peak below the threshold is the signal of the
$\bar{K} N N$ bound state, while the peak above the threshold
originates from the quasi-elastic scattering of the kaon in the first
collision emitting a fast nucleon, in processes of the type $K^{-} n
\to K^{-} n_{\rm escape}$ or $K^{-} p \to \bar{K}^{0} n_{\rm escape}$.

This paper is organized as follows. In Sec.~\ref{sec:2} we develop our
formulation to obtain the cross section of the ${}^{3} \text{He} (
K^{-}, \, \Lambda p ) n$ reaction, describing the details that allow
us to obtain the scattering amplitude for the uncorrelated $\Lambda
(1405) p$ mechanism and for the $\bar{K} N N$ bound state one. Next,
in Sec.~\ref{sec:3} we show our results and discuss the origin of the
peak structure observed in the in-flight ${}^{3} \text{He} ( K^{-}, \,
\Lambda p ) n$ reaction in the J-PARC experiment.  Section~\ref{sec:4}
is devoted to the conclusions of this study.

\section{Formulation}
\label{sec:2}

In this section we formulate the cross section and scattering
amplitude of the in-flight ${}^{3} \text{He} ( K^{-}, \, \Lambda p )
n$ reaction. After showing the expression of the cross section in
Sec.~\ref{sec:2-1}, we construct the scattering amplitude of the
reaction in Sec.~\ref{sec:2-2} and Sec.~\ref{sec:2-3}.  In
Sec.~\ref{sec:2-2} we consider the case of an uncorrelated $\Lambda
(1405) p$ system, i.e., the $\Lambda (1405) p$ system is generated
without binding after emission of a fast neutron, and in
Sec.~\ref{sec:2-3} we take into account the multiple scattering of
$\bar{K}$ between two nucleons to generate a $\bar{K} N N$ quasi-bound
state.

According to the experimental condition, we concentrate on the
three-nucleon absorption of $K^{-}$, i.e., we do not allow a spectator
nucleon.  Throughout this study, we take the global center-of-mass
frame when we calculate the phase space for the cross section, while
we evaluate the scattering amplitude in the ${}^{3} \text{He}$ rest
frame so as to omit the center-of-mass momentum of ${}^{3} \text{He}$
in the wave function.  Throughout this work the physical masses for
the hadrons are used, except in the evaluation of the $\HeT$ wave
function and in the kaon propagators of the multiple scattering
$\bar{K} N N$ amplitude.

\subsection{Cross section of the $\HeT ( K^{-} , \, \Lambda p ) n$ reaction}
\label{sec:2-1}

First we formulate the cross section for the $K^{-} ( k ) \HeT ( P )
\to \Lambda ( p_{\Lambda}^{\prime} ) p ( p_{p}^{\prime} ) n (
p_{n}^{\prime} ) $ reaction, where the momenta of these particles are
shown in parentheses.  Since we are interested in the $\Lambda p$
spectrum as a function of its invariant mass $M_{\Lambda p}$, we fix
the final-state phase space with the invariant mass $M_{\Lambda p}$,
the solid angle for the neutron momentum in the global center-of-mass
frame $\Omega _{n}$, and the solid angle for $\Lambda$ in the
$\Lambda$-$p$ rest frame $\Omega _{\Lambda}^{\ast}$.  With these
quantities, the differential cross section can be expressed
as~\cite{Agashe:2014kda}:
\begin{equation}
  \frac{d^{2} \sigma}{d M_{\Lambda p} d \cos \theta _{n}^{\rm cm}}
  = \frac{M_{{}^{3} \text{He}} m_{\Lambda} m_{p} m_{n}}
  {( 2 \pi )^{4} 2 p_{\rm cm} E_{\rm cm}^{2}}
  p_{n}^{\prime} p_{\Lambda}^{\ast}
  \, \int d \Omega _{\Lambda}^{\ast} 
  \overline{\sum _{\lambda}} \sum _{\lambda ^{\prime}}
  | \mathcal{T} | ^{2} ,
  \label{eq:ds}
\end{equation}
where we have performed the integral with respect to the azimuthal
angle for the neutron momentum, which is irrelevant to the present
formulation.  In the expression, $\theta _{n}^{\rm cm}$ is the neutron
scattering angle in the global center-of-mass frame, $M_{{}^{3}
  \text{He}}$, $m_{\Lambda}$, $m_{p}$, and $m_{n}$ are the masses of
${}^{3} \text{He}$, $\Lambda$, proton and neutron, respectively, and
$p_{\rm cm}$ and $E_{\rm cm}$ are the center-of-mass momentum and
energy for the initial state:
\begin{equation}
  p_{\rm cm} \equiv \frac{\lambda ^{1/2} ( E_{\rm cm}^{2} , \, m_{K^{-}}^{2}, \,
    M_{{}^{3} \text{He}}^{2})}{2 E_{\rm cm}} ,
  \quad
  E_{\rm cm} \equiv \sqrt{ ( k + P )^{2} }
  = \sqrt{m_{K^{-}}^{2} + M_{\HeT}^{2}
    + 2 \omega _{K^{-}} ( \bm{k} ) M_{\HeT}} ,
\end{equation}
with the $K^{-}$ mass $m_{K^{-}}$, the \Kaellen function $\lambda ( x
, \, y , \, z ) = x^{2} + y^{2} + z^{2} - 2 x y - 2 y z - 2 z x$, the
initial kaon momentum in the laboratory frame $k^{\mu}$, and $\omega
_{K^{-}} ( \bm{k} ) \equiv \sqrt{\bm{k}^{2} + m_{K^{-}}^{2}}$.  The
momenta $p_{n}^{\prime}$ and $p_{\Lambda}^{\ast}$ correspond to that of the neutron
in the global center-of-mass frame and that of the $\Lambda$ in the
$\Lambda$-$p$ rest frame, respectively, evaluated as
\begin{equation}
  p_{n}^{\prime} \equiv \frac{\lambda ^{1/2} ( E_{\rm cm}^{2} , \,
    M_{\Lambda p}^{2} , \, m_{n}^{2})}{2 E_{\rm cm}} ,
  \quad
  p_{\Lambda}^{\ast} \equiv \frac{\lambda ^{1/2} ( M_{\Lambda p}^{2} , \,
    m_{\Lambda}^{2} , \, m_{p}^{2})}{2 M_{\Lambda p}} .
\end{equation}
By means of the summation symbols in Eq.~\eqref{eq:ds}, we perform the
average and sum of the squared scattering amplitude, $| \mathcal{T}
|^{2}$, for the polarizations of the initial- and final-state
particles, respectively.

From the double differential cross section $d^{2} \sigma / d M_{\Lambda p} d
\cos \theta _{n}^{\rm cm}$, we can evaluate the mass spectrum $d
\sigma / d M_{\Lambda p}$ and the differential cross section with
respect to the neutron angle $d \sigma / d \cos \theta _{n}^{\rm cm}$
by performing the integral with respect to $\cos \theta _{n}^{\rm
  cm}$ and $M_{\Lambda p}$, respectively:
\begin{equation}
  \frac{d \sigma}{d M_{\Lambda p}} = \int _{-1}^{1} d \cos \theta _{n}^{\rm cm}
  \frac{d^{2} \sigma}{d M_{\Lambda p} d \cos \theta _{n}^{\rm cm}} ,
  \quad
  \frac{d \sigma}{d \cos \theta _{n}^{\rm cm}} =
  \int _{M_{\rm min}}^{M_{\rm max}} d M_{\Lambda p}
  \frac{d^{2} \sigma}{d M_{\Lambda p} d \cos \theta _{n}^{\rm cm}} ,
\end{equation}
where $M_{\rm min}$ and $M_{\rm max}$ are the lower and upper bounds
of the invariant mass $M_{\Lambda p}$, respectively, which become
$M_{\rm min} \sim 2.1 \gev$ and $M_{\rm max} \sim 2.9 \gev$ for an
initial kaon momentum in the laboratory frame of $k_{\rm lab} = 1 \gev
/c$, as that employed in the J-PARC E15
experiment~\cite{Sada:2016vkt}.  In this study, however, we restrict
those values to $M_{\rm min} = 2.2 \gev$ and $M_{\rm max} = 2.6 \gev$,
since we are interested only in the physics leading to a peak around
the $\bar{K} N N$ threshold and we are ignoring other contributions
which only play a background role in this region.  We can also
evaluate the total cross section as
\begin{equation}
  \sigma =   \int _{M_{\rm min}}^{M_{\rm max}} d M_{\Lambda p}
  \int _{-1}^{1} d \cos \theta _{n}^{\rm cm}
  \frac{d^{2} \sigma}{d M_{\Lambda p} d \cos \theta _{n}^{\rm cm}} .
\end{equation}

\subsection{Scattering amplitude: generating an uncorrelated
  $\Lambda (1405) p$ system}
\label{sec:2-2}

\begin{figure}[!t]
  \centering
  \PsfigII{0.25}{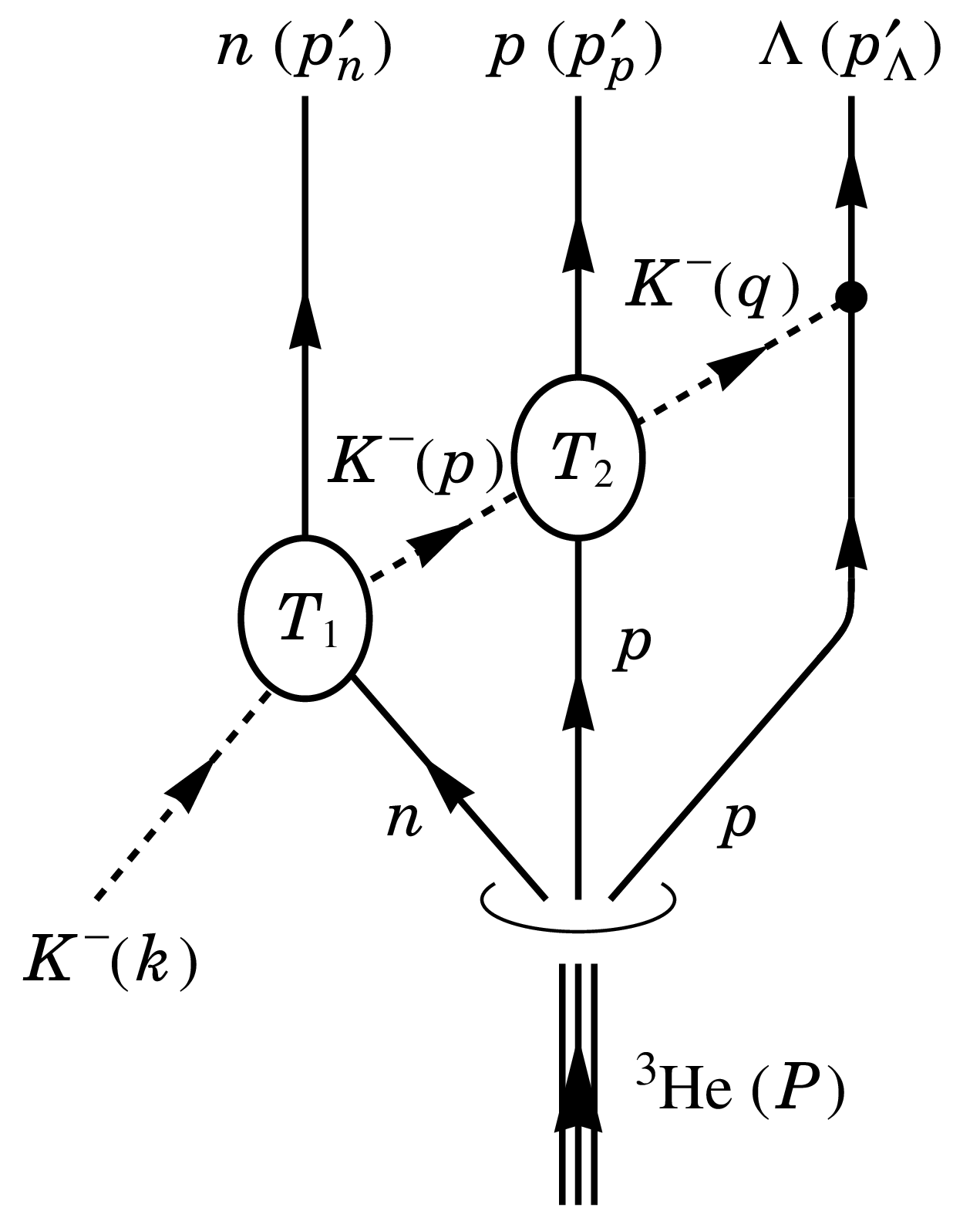} ~ \PsfigII{0.25}{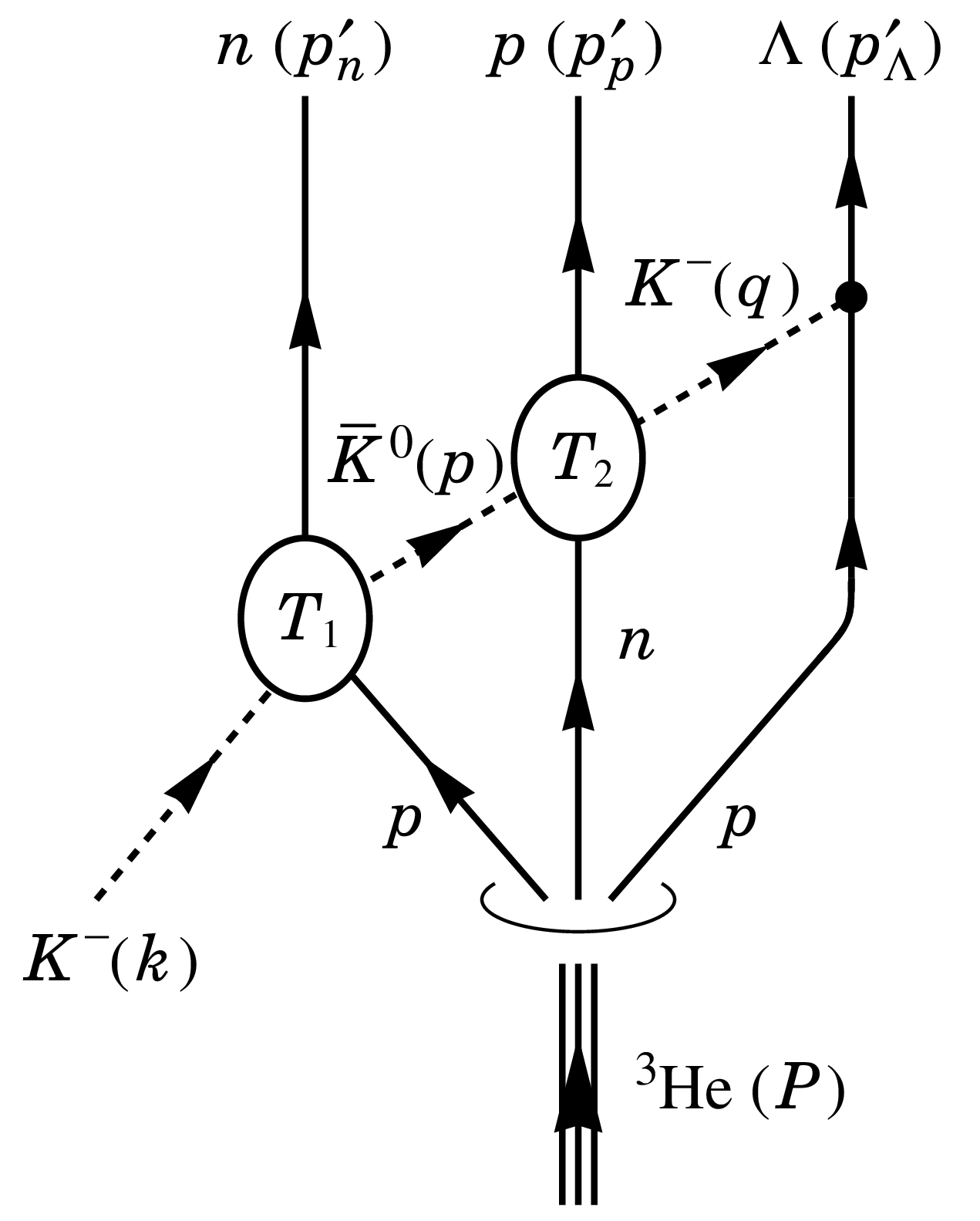} 
  \caption{Two Feynman diagrams most relevant to the three-nucleon
    absorption of $K^{-}$ via an uncorrelated $\Lambda (1405) p$
    system.  We also show momenta of particles in parentheses.}
  \label{fig:3NA}
\end{figure}

Next we construct the scattering amplitude of the ${}^{3} \text{He} (
K^{-}, \, \Lambda p ) n$ reaction with three-nucleon absorption of
$K^{-}$.  In this subsection we consider a case that an uncorrelated
$\Lambda (1405) p$ system, without binding, is generated after the
emission of a fast neutron.  This process has a possibility of making
a peak structure around the $K^{-} p p$ threshold in the $\Lambda p$
invariant mass spectrum, since the $\Lambda (1405)$ resonance appears
below the $K^{-} p$ threshold.  In Fig.~\ref{fig:3NA} we show the two
diagrams which implicitly contain the $\Lambda (1405)$ resonance in
the intermediate state.  In these diagrams, a fast neutron is emitted
after the first collision of the $K^{-}$ with a nucleon ($T_{1}$).
Then, the second collision of the $K^{-} p$ or the $\bar{K}^{0} n$
($T_{2}$) is enhanced at the energy of the $\Lambda (1405)$ resonance,
producing an enhancement around the $K^{-} p p$ threshold.  Finally,
the rescattered $K^{-}$ is absorbed into a proton to become a
$\Lambda$ particle.  An important point is that one expects a kaon,
rather than another meson like the $\eta$ or the pion, to be absorbed
by the last nucleon because 1) the propagating kaon after the first
collision is almost on its mass shell, hence the amplitude $T_{2}$
covers the region of the $\Lambda (1405)$ resonance, 2) the $\Lambda
(1405)$ is expected to be a $\bar{K} N (I=0)$ bound
state~\cite{Hall:2014uca, Sekihara:2014kya, Kamiya:2015aea}, hence the
process is dominated by the isospin $I=0$ component of the $\bar{K} N$
interaction, and 3) the coupling strength of the $K^{-} p \Lambda$
vertex is strong while that of $\eta p p$ is very weak in flavor SU(3)
symmetry~\cite{Sekihara:2009yk, Sekihara:2012wj}.\footnote{This latter
  vertex would appear in a mechanism where $T_{2}$ accounts for the
  $\bar{K} N \to \eta \Lambda$ scattering, followed by the $\eta N N$
  Yukawa vertex.}  Therefore, the most relevant diagrams for the
reaction are the two shown in Fig.~\ref{fig:3NA}.  The contribution
from the uncorrelated $\Lambda (1405) p$ system will be observed as a
peak in the $\Lambda p$ invariant mass spectrum.

Due to the antisymmetrization for the two protons in $\HeT$, we have
two contributions from each diagram in Fig.~\ref{fig:3NA}.  Therefore,
the scattering amplitude of the reaction $\mathcal{T}$ can be
expressed as
\begin{equation}
  \mathcal{T} =
  \mathcal{T}_{1} + \mathcal{T}_{2} + \mathcal{T}_{3} + \mathcal{T}_{4} .
\end{equation}
The antisymmetrized $\HeT$ wave function is given in
Appendix~\ref{app:1}, and $\mathcal{T}_{1, 2, 3, 4}$ come from the
first, second, third, and fourth terms of the $\HeT$ wave function in
Eq.~\eqref{eqA:WFspin}, respectively.\footnote{The last two terms in
  Eq.~\eqref{eqA:WFspin} have the neutron in the third place and
  induce a $\bar{K}^{0} p \to \bar{K}^{0} p$ interaction in $T_{2}$,
  which has $I = 1$ and hence negligible compared to the $\Lambda
  (1405)$ excitation in $I = 0$.}  In the following we give the
explicit form of each amplitude $\mathcal{T}_{i}$.

Let us first fix the amplitude $\mathcal{T}_{1}$, which comes from the
left diagram in Fig.~\ref{fig:3NA}.  The momenta of the three
nucleons in $\HeT$ are denoted as $p_{1}^{\mu}$, $p_{2}^{\mu}$, $p_{3}^{\mu}$ from
left to right.  Thus, we have
\begin{equation}
  p_{1}^{\mu} = ( p + p_{n}^{\prime} - k )^{\mu} ,
  \quad 
  p_{2}^{\mu} = ( q + p_{p}^{\prime} - p )^{\mu} , 
  \quad
  p_{3}^{\mu} = ( p_{\Lambda}^{\prime} - q )^{\mu} .
\end{equation}
From these momenta, we can construct the momenta in Jacobi
coordinates as in Appendix~\ref{app:1}, and here we show only the
relevant expressions to the present formulation:
\begin{equation}
  \begin{split}
    & \bm{P} = 
    \bm{p}_{1} + \bm{p}_{2} + \bm{p}_{3} ,
    \\
    & \bm{p}_{\lambda} = 
    \frac{2 \bm{p}_{1} - \bm{p}_{2} - \bm{p}_{3}}{3}
    = \bm{p} + \bm{p}_{n}^{\prime} - \bm{k}
    - \frac{1}{3} \bm{P},
    \quad
    \bm{p}_{\rho} = 
    \frac{\bm{p}_{3} - \bm{p}_{2}}{2}
    = \frac{\bm{p} - 2 \bm{q} + \bm{p}_{\Lambda}^{\prime}
    - \bm{p}_{p}^{\prime}}{2} .
  \end{split}
\end{equation}
Then, by using a scheme similar to that developed in
Refs.~\cite{Jido:2009jf, Jido:2010rx, YamagataSekihara:2012yv,
  Jido:2012cy} and the $\HeT$ wave function summarized in
Appendix~\ref{app:1}, we can evaluate the scattering amplitude
$\mathcal{T}_{1}$ as
\begin{align}
  - i \mathcal{T}_{1} = & \int \frac{d^{3} q}{( 2 \pi )^{3}}
  \frac{i}{(q^{0})^{2} -  \omega _{K^{-}} ( \bm{q} )^{2}}
  \int \frac{d^{3} p}{( 2 \pi )^{3}}
  \frac{i}{(p^{0})^{2} - \omega _{K^{-}} ( \bm{p} )^{2} + i m_{K^{-}} \Gamma _{K}}
  \, \tilde{\Psi} ( p_{\lambda} , \, p_{\rho} )
  \notag \\ & \times 
  \left [ - i \chi _{p}^{\dagger} T_{2}^{(K^{-} p \to K^{-} p)} ( w_{2} )
    \chi _{\uparrow} \right ]
  \left [ - i \chi _{n}^{\dagger}
    T_{1}^{(K^{-} n \to K^{-} n)} ( w_{1} , \, \cos \theta _{1} ) \chi \right ]
  \notag \\
  & \times
  \left [ \tilde{V} \mathcal{F} ( \bm{q} )
    \bm{q}  \left ( \chi _{\Lambda}^{\dagger}
    \bm{\sigma} \chi _{\downarrow} \right ) \right ]
  ,
\end{align}
where $\Gamma _{K}$ is the kaon absorption width by two nucleons in
the $\bar{K} N N$ system, whose value is fixed to be $\Gamma _{K} = 15
\mev$ so as to reproduce the kaon absorption width of the $\bar{K} N
N$ bound state in the fixed center approximation~\cite{Bayar:2012hn}
(see Fig.~\ref{fig:T_FCA} and related discussions below).  We note
that the exchanged kaon after the secondary scattering ($q^{\mu}$)
goes highly off its mass shell in the present kinematics, so the term
$i m_{K^{-}} \Gamma _{K}$ is unnecessary in the denominator of the
corresponding propagator.  The energies of the intermediate kaons are
fixed in two ways: one employs the Watson
approach~\cite{Watson:1953zz} and the other one relies on the
truncated Faddeev approach~\cite{Miyagawa:2012xz}, to which we refer
as options A and B, respectively.  Namely, in option A we
have~\cite{Jido:2012cy}
\begin{equation}
  q^{0} = p_{\Lambda}^{\prime \, 0}
  - \left ( m_{p} - \frac{B_{\HeT}}{3} \right ) ,
\end{equation}
\begin{equation}
  p^{0} = q^{0} + p_{p}^{\prime \, 0}
  - \left ( m_{p} - \frac{B_{\HeT}}{3} \right ) 
  = p_{\Lambda}^{\prime \, 0} + p_{p}^{\prime \, 0}
  - 2 \left ( m_{p} - \frac{B_{\HeT}}{3} \right ) ,
\end{equation}
with $B_{\HeT} = 7.7 \mev$ being the $\HeT$ binding energy, while in option
B we have~\cite{Miyagawa:2012xz}
\begin{equation}
  q^{0} = p_{\Lambda}^{\prime \, 0}
  - \mathcal{E}_{p} ( \bm{p}_{\Lambda}^{\prime} - \bm{q} ) ,
\end{equation}
\begin{align}
  p^{0} = & q^{0} + p_{p}^{\prime \, 0}
  - \mathcal{E}_{p} ( \bm{q} + \bm{p}_{p}^{\prime} - \bm{p} )
  = p_{\Lambda}^{\prime \, 0} + p_{p}^{\prime \, 0}
  - \mathcal{E}_{p} ( \bm{p}_{\Lambda}^{\prime} - \bm{q} )
  - \mathcal{E}_{p} ( \bm{q} + \bm{p}_{p}^{\prime} - \bm{p} ) ,
\end{align}
where $\mathcal{E}_{p} ( \bm{q} ) \equiv m_{p} + \bm{q}^{2} / (2
m_{p})$.  The Watson approach contains more
contributions from diagrams of the rescattering of nucleons via the $N
N$ interaction, while the truncated Faddeev approach can give a
correct threshold behavior (see Ref.~\cite{Jido:2012cy} for 
details).

The spinors $\chi$, $\chi _{\Lambda}$, $\chi _{p}$, and $\chi _{n}$
stand for initial-state $\HeT$, and final-state $\Lambda$, proton, and
neutron, respectively, all of which being either $\chi_{\uparrow} = (
1 , \, 0 )^{t}$ or $\chi_{\downarrow} = ( 0 , \, 1 )^{t}$.  Since we
assume that the spin direction of $\HeT$ equals that of the bound
neutron, we take the same spinor for both of them.  The $\HeT$ wave
function $\tilde{\Psi}$ is evaluated with the harmonic oscillator
potential, and its explicit form is given in
Appendix~\ref{app:1}.\footnote{We would have a factor $1/\sqrt{6}$
  from the $\HeT$ wave function as in Eq.~\eqref{eqA:WFspin}, but this
  factor will be compensated in the cross section by the identical
  contributions of six diagrams of different topology.  These
  correspond to having the first scattering, $T_{1}$, in either of the
  three nucleons and the second scattering, $T_{2}$, in either of the
  remaining two nucleons.  The final states have the triplet $n p
  \Lambda$ produced in different order and these contributions add
  incoherently in the cross section.}

The Yukawa $K^{-} p \Lambda$ vertex gives rise to Pauli matrices
$\bm{\sigma}$ and a coupling constant $\tilde{V}$
\begin{equation}
  \tilde{V} 
  = \alpha \frac{D + F}{2 f}
  + \beta \frac{D - F}{2 f} ,
  \quad
  \alpha = \frac{2}{\sqrt{3}} ,
  \quad
  \beta = - \frac{1}{\sqrt{3}} ,
  \label{eq:Vmbb}
\end{equation}
where $f$ is the meson decay constant, taken to be $f = 93 \mev$,
while $D = 0.795$ and $F = 0.465$ are adjusted to the weak decay of
baryons.  We also introduce a form factor 
\begin{equation}
  \mathcal{F} ( \bm{q} ) = \frac{\Lambda ^{2}}{\Lambda ^{2} + \bm{q}^{2}} ,
  \label{eq:FF_vertex}
\end{equation}
for this vertex.  We take a typical cutoff value $\Lambda = 0.8 \gev$,
but the cutoff dependence of the cross section will be discussed later
on.

The $K^{-} n \to K^{-} n$ scattering amplitude, $T_{1}^{(K^{-} n \to
  K^{-} n)}$, is a function of the center-of-mass energy for the
initial kaon-bound neutron system, $w_{1}$, and the scattering angle
in their center-of-mass frame $\theta _{1}$, both of which are
evaluated by neglecting the Fermi motion of the bound neutron.  As a
result, we have
\begin{equation}
  w_{1} = \sqrt{( k + p_{1} )^{2}}
  \approx \sqrt{\left ( k^{0} + m_{n} - B_{\HeT} / 3 \right ) ^{2}
    - \bm{k}^{2}} ,
  \label{eq:w1}
\end{equation}
\begin{equation}
  \cos \theta _{1} = \frac{m_{n}^{2} + m_{K^{-}}^{2}
    - 2 \omega _{K^{-}} ( p_{K^{-}} ( w_{1} ) ) E_{n} ( w_{1} )
    - (p_{n}^{\prime} - k)^{2}}{2 p_{K^{-}} ( w_{1} )^{2}} ,
  \label{eq:costheta1}
\end{equation}
where
\begin{equation}
  p_{K^{-}} ( w_{1} ) \equiv \frac{\lambda ^{1/2} ( w_{1}^{2}, \,
  m_{K^{-}}^{2} , \, m_{n}^{2})}{2 w_{1}} ,
  \quad
  E_{n} ( w_{1} ) \equiv \frac{w_{1}^{2} + m_{n}^{2} - m_{K^{-}}^{2}}{2 w_{1}} .
\end{equation}
We note that the value of right-hand side in Eq.~\eqref{eq:costheta1}
may become larger than $1$ or smaller than $-1$ because the bound
nucleons actually have an energy and momentum distribution different
to the free one.  In such a case we take $\cos \theta _{1} = 1$ or
$-1$, respectively.  Now that $w_{1}$ and $\theta _{1}$ are fixed by
the momenta of the initial- and final-state particles, we can put
$T_{1}^{(K^{-} n \to K^{-} n)}$ outside of the integral.  In addition,
for this amplitude $T_{1}$ we neglect the spin flip contribution, and
hence we can factorize the spinor part $\chi _{n}^{\dagger} \chi$.
The amplitude $T_{1}^{(K^{-} n \to K^{-} n)}$ is evaluated
phenomenologically in Appendix~\ref{app:2}.

The $K^{-} p \to K^{-} p$ scattering amplitude,
$T_{2}^{(K^{-} p \to K^{-} p)}$, is a function of the center-of-mass
energy for the exchanged kaon ($q^{\mu}$) and final-state proton,
$w_{2}$:
\begin{equation}
  w_{2} = \sqrt{( q + p_{p}^{\prime} )^{2}}
    = \sqrt{( q^{0} + p_{p}^{\prime \, 0} )^{2}
      - | \bm{q} + \bm{p}_{p}^{\prime} |^{2} } .
\end{equation}
Since the relevant energies to our study are those near the $K^{-} p$
threshold, we only consider the $s$-wave part of the amplitude
$T_{2}^{(K^{-} p \to K^{-} p)}$.  We calculate this $K^{-} p \to K^{-}
p$ amplitude in the so-called chiral unitary
approach~\cite{Kaiser:1995eg, Oset:1997it, Oller:2000fj, Jido:2003cb},
where the $\Lambda (1405)$ resonance is dynamically generated from the
meson-baryon degrees of freedom.  For the chiral unitary amplitude we
take into account ten channels: $K^{-} p$, $\bar{K}^{0} n$, $\pi ^{0}
\Lambda$, $\pi ^{0} \Sigma ^{0}$, $\pi ^{+} \Sigma ^{-}$, $\pi ^{-}
\Sigma ^{+}$, $\eta \Lambda$, $\eta \Sigma ^{0}$, $K^{0} \Xi ^{0}$,
and $K^{+} \Xi ^{-}$.  The formulation of the amplitude $T_{2}$ in the
chiral unitary approach is summarized in Appendix~\ref{app:3}.  An
important point is that in this amplitude we take into account the
kaon absorption by two nucleons in the $\bar{K} N N$ system
effectively via the inclusion of a width $\Gamma _{K}$ in the kaon
propagator [see Fig.~\ref{fig:app3}(a) in Appendix~\ref{app:3}].

As a consequence, the explicit form of $\mathcal{T}_{1}$ finally
becomes
\begin{align}
  \mathcal{T}_{1} = & i \left ( \chi _{n}^{\dagger} \chi \right )
  \left ( \chi _{p}^{\dagger} \chi _{\uparrow} \right ) \times
  T_{1}^{(K^{-} n \to K^{-} n)} ( w_{1}, \, \cos \theta _{1} )
  \tilde{V} 
  \int \frac{d^{3} q}{( 2 \pi )^{3}} 
  \frac{\mathcal{F} ( \bm{q} )
    \bm{q}  \left ( \chi _{\Lambda}^{\dagger} \bm{\sigma}
    \chi _{\downarrow} \right )}{(q^{0})^{2} - \omega _{K^{-}} ( \bm{q} )^{2}}
  \notag \\ & \times
  T_{2}^{(K^{-} p \to K^{-} p)} ( w_{2} ) 
  \int \frac{d^{3} p}{( 2 \pi )^{3}}
  \frac{1}{(p^{0})^{2} - \omega _{K^{-}} ( \bm{p} )^{2} + i m_{K^{-}} \Gamma _{K}}
  \, 
  \tilde{\Psi} ( p_{\lambda} , \, p_{\rho} ) .
  \label{eq:mathT1}
\end{align}

In a similar manner, we can write the formulas of the other scattering
amplitudes.  Here we note that, although one needs to antisymmetrize the
momentum and spin of the nucleons in $\HeT$, the wave function
$\tilde{\Psi} ( p_{\lambda} , \, p_{\rho} )$ in
Eq.~\eqref{eqA:tildePsi} is unchanged for the exchange of momenta
$\bm{p}_{i} \leftrightarrow \bm{p}_{j}$ ($i$, $j = 1$, $2$, and $3$),
since the global argument of the Gaussian functions in $\tilde{\Psi} (
p_{\lambda} , \, p_{\rho} )$ reduces to $\left ( \bm{P}^{2}/3 - \sum
_{i=1}^{3} \bm{p}_{i}^{2} \right ) / (2 m_{N} \omega
_{\lambda})$. Therefore, we have to consider the antisymmetrization
of the spin variables only, and we have
\begin{align}
  \mathcal{T}_{2} = & - i \left ( \chi _{n}^{\dagger} \chi \right )
  \left ( \chi _{p}^{\dagger} \chi _{\downarrow} \right ) \,
  T_{1}^{(K^{-} n \to K^{-} n)} ( w_{1}, \, \cos \theta _{1} )
  \tilde{V} 
  \int \frac{d^{3} q}{( 2 \pi )^{3}} 
  \frac{\mathcal{F} ( \bm{q} )
    \bm{q}  \left ( \chi _{\Lambda}^{\dagger} \bm{\sigma}
    \chi _{\uparrow} \right )}{(q^{0})^{2} - \omega _{K^{-}} ( \bm{q} )^{2}}
  \notag \\ & \times
  T_{2}^{(K^{-} p \to K^{-} p)} ( w_{2} ) 
  \int \frac{d^{3} p}{( 2 \pi )^{3}}
  \frac{1}{(p^{0})^{2} - \omega _{K^{-}} ( \bm{p} )^{2} + i m_{K^{-}} \Gamma _{K}}
  \, 
  \tilde{\Psi} ( p_{\lambda} , \, p_{\rho} ) ,
\end{align}
\begin{align}
  \mathcal{T}_{3} = & - i \left ( \chi _{n}^{\dagger} \chi _{\uparrow} \right )
  \left ( \chi _{p}^{\dagger} \chi \right ) \,
  T_{1}^{(K^{-} p \to \bar{K}^{0} n)} ( w_{1}^{\prime}, \, \cos \theta _{1}^{\prime} )
  \tilde{V} 
  \int \frac{d^{3} q}{( 2 \pi )^{3}} 
  \frac{\mathcal{F} ( \bm{q} )
    \bm{q}  \left ( \chi _{\Lambda}^{\dagger} \bm{\sigma}
    \chi _{\downarrow} \right )}{(q^{0})^{2} - \omega _{K^{-}} ( \bm{q} )^{2}}
  \notag \\ & \times
  T_{2}^{(\bar{K}^{0} n \to K^{-} p)} ( w_{2} ) 
  \int \frac{d^{3} p}{( 2 \pi )^{3}}
  \frac{1}{(p^{\prime \, 0})^{2} - \omega _{\bar{K}^{0}} ( \bm{p} )^{2} + i m_{\bar{K}^{0}}
    \Gamma _{K}}
  \, 
  \tilde{\Psi} ( p_{\lambda} , \, p_{\rho} ) ,
\end{align}
\begin{align}
  \mathcal{T}_{4} = & i \left ( \chi _{n}^{\dagger} \chi _{\downarrow} \right )
  \left ( \chi _{p}^{\dagger} \chi \right ) \,
  T_{1}^{(K^{-} p \to \bar{K}^{0} n)} ( w_{1}^{\prime}, \, \cos \theta _{1}^{\prime} )
  \tilde{V} 
  \int \frac{d^{3} q}{( 2 \pi )^{3}} 
  \frac{\mathcal{F} ( \bm{q} )
    \bm{q}  \left ( \chi _{\Lambda}^{\dagger} \bm{\sigma}
    \chi _{\uparrow} \right )}{(q^{0})^{2} - \omega _{K^{-}} ( \bm{q} )^{2}}
  \notag \\ & \times
  T_{2}^{(\bar{K}^{0} n \to K^{-} p)} ( w_{2} ) 
  \int \frac{d^{3} p}{( 2 \pi )^{3}}
  \frac{1}{(p^{\prime \, 0})^{2} - \omega _{\bar{K}^{0}} ( \bm{p} )^{2} + i m_{\bar{K}^{0}}
    \Gamma _{K}}
  \, 
  \tilde{\Psi} ( p_{\lambda} , \, p_{\rho} ) .
\end{align}
Here $\omega _{\bar{K}^{0}} ( \bm{p} ) \equiv \sqrt{\bm{p}^{2} +
  m_{\bar{K}^{0}}^{2}}$ with the $\bar{K}^{0}$ mass $m_{\bar{K}^{0}}$
and $w_{1}^{\prime}$ and $\cos \theta _{1}^{\prime}$ are fixed in the
same manner as in Eqs.~\eqref{eq:w1} and \eqref{eq:costheta1},
respectively, but for the $K^{-} p \to \bar{K}^{0} n$ reaction instead
of the $K^{-} n \to K^{-} n$ one.  The energy $p^{\prime \, 0}$ is fixed as
\begin{align}
  p^{\prime \, 0}
  =
  \begin{cases}
    \displaystyle p_{\Lambda}^{\prime \, 0} + p_{p}^{\prime \, 0}
    - \left ( m_{p} + m_{n} - \frac{2}{3} B_{\HeT} \right )
    & \text{ in option A} ,
    \\ 
    p_{\Lambda}^{\prime \, 0} + p_{p}^{\prime \, 0}
    - \mathcal{E}_{p} ( \bm{p}_{\Lambda}^{\prime} - \bm{q} )
    - \mathcal{E}_{n} ( \bm{q} + \bm{p}_{p}^{\prime} - \bm{p} )
    & \text{ in option B} ,
  \end{cases}
\end{align}
with $\mathcal{E}_{n} ( \bm{q} ) \equiv m_{n} + \bm{q}^{2} / (2
m_{n})$, where $p^{\prime \, 0}$ is different from $p^{0}$ as the
former contains the neutron mass or energy.  The amplitudes
$T_{1}^{(K^{-} n \to K^{-} n)}$ and $T_{1}^{(K^{-} p \to \bar{K}^{0}
  n)}$ are evaluated phenomenologically in Appendix~\ref{app:2}, while
the amplitudes $T_{2}^{(K^{-} p \to K^{-} p)}$ and
$T_{2}^{(\bar{K}^{0} n \to K^{-} p)}$ are taken from a chiral unitary approach in $s$ wave, as described in Appendix~\ref{app:3}.

\subsection{Scattering amplitude: generating a $\bar{K} N N$ quasi-bound state}
\label{sec:2-3}

\begin{figure}[b]
  \centering
  \PsfigII{0.225}{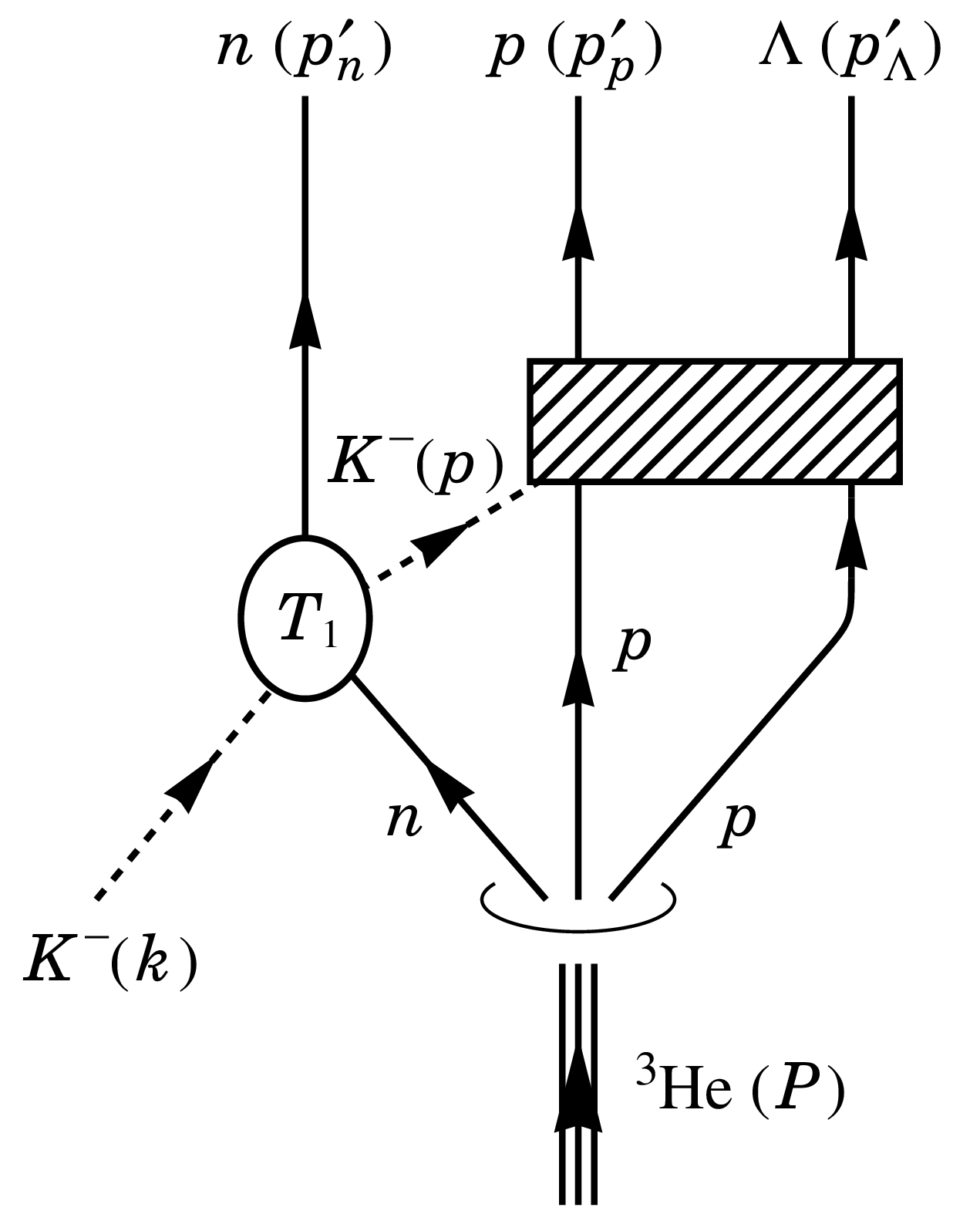} ~ \PsfigII{0.225}{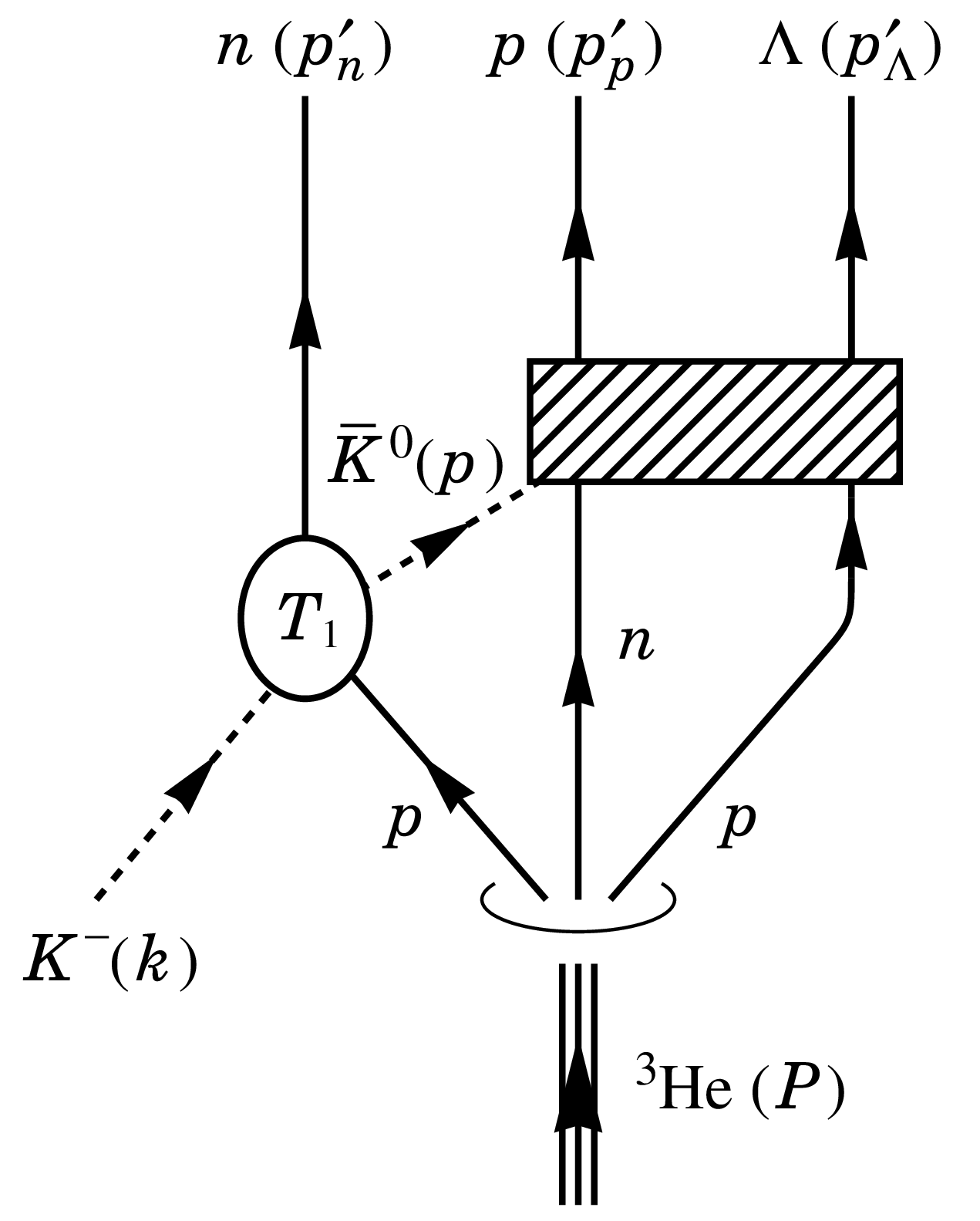} 
  ~ \PsfigII{0.225}{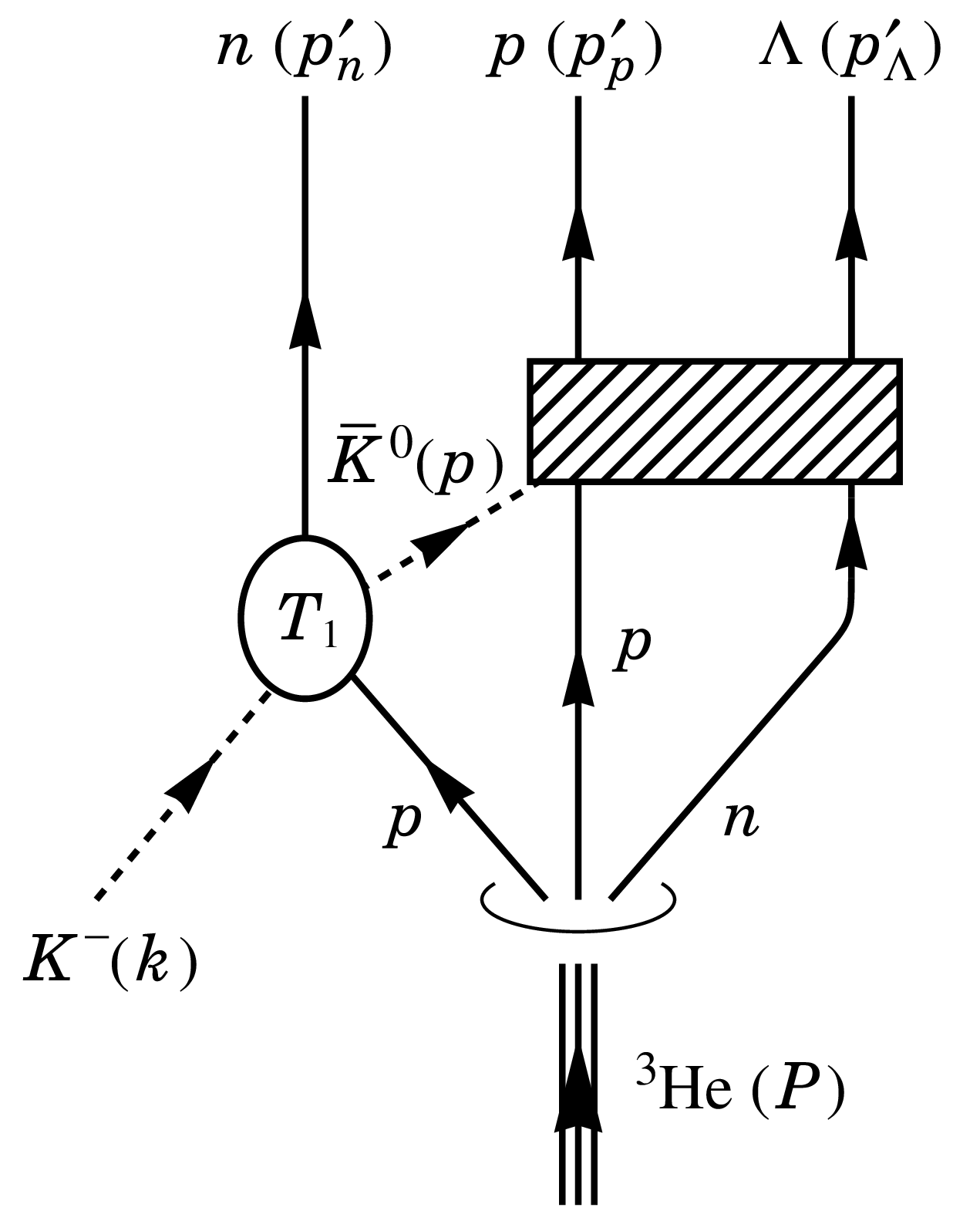} 
  \caption{The three most relevant Feynman diagrams depicting the
    three-nucleon absorption of a $K^{-}$ implementing the multiple
    kaon scattering between two nucleons, which is represented by the
    shaded rectangles (see Fig.~\ref{fig:FCA}).}
  \label{fig:3NA_FCA}
\end{figure}

In this subsection we consider the case of the formation of a $\bar{K} N N$ quasi-bound
state in the ${}^{3} \text{He} ( K^{-}, \, \Lambda p ) n$
reaction, which would be the origin of the peak structure seen in the
J-PARC E15 experiment.  The $\bar{K} N N$ quasi-bound state is
generated as the multiple scattering of the kaon between two nucleons
after emission of a fast neutron in the reaction.  The most relevant
diagrams are shown in Fig.~\ref{fig:3NA_FCA}.

Taking into account the antisymmetrization for three nucleons, we have
six contributions to the scattering amplitude of the reaction:
\begin{equation}
  \mathcal{T} =
  \mathcal{T}_{1} + \mathcal{T}_{2} + \mathcal{T}_{3} + \mathcal{T}_{4}
  + \mathcal{T}_{5} + \mathcal{T}_{6} .
\end{equation}
The amplitudes $\mathcal{T}_{1, 2, 3, 4, 5, 6}$ come from the first,
second, third, fourth, fifth, and sixth terms of the $\HeT$ wave
function in Eq.~\eqref{eqA:WFspin} in Appendix~\ref{app:1},
respectively.

\begin{figure}[t]
  \centering
  \PsfigII{0.225}{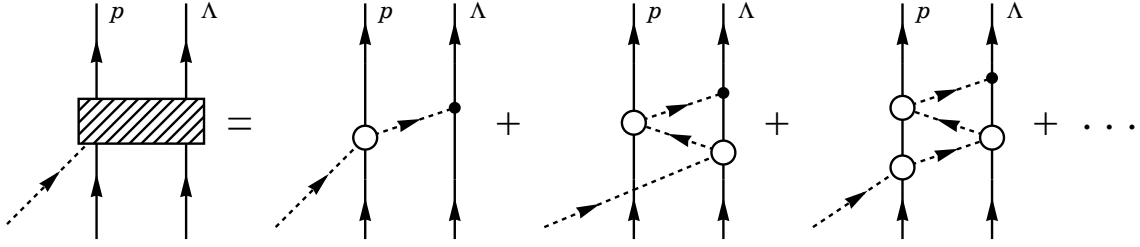} 
  \caption{Diagrammatic equation for the kaon absorption amplitudes
    after multiple scattering.  They contain multiple scattering
    amplitudes, which are evaluated in the fixed center approximation.
    The unspecified dashed and solid lines are $\bar{K}$ and $N$,
    respectively.  The open circles are the $\bar{K} N \to \bar{K} N$
    scattering amplitude evaluated in the chiral unitary approach and
    the dots represent the $\bar{K} N \Lambda$ vertex.}
  \label{fig:FCA}
\end{figure}

Let us consider the first term $\mathcal{T}_{1}$.  We can use the same
form as in Eq.~\eqref{eq:mathT1}, keeping the scattering amplitude of
the first step, $T_{1}^{(K^{-} n \to K^{-} n)}$, the $K^{-}$
propagator of $p^{\mu}$, and the $\HeT$ wave function.  Then, the most
important part of the scattering amplitude, i.e. the part where the
$\bar{K} N N$ quasi-bound state is generated and the kaon is absorbed,
remains to be implemented.  This is represented by the diagrams shown
in Fig.~\ref{fig:FCA}, which are calculated as follows.  First, we do
not consider spin flips during the multiple scattering since the
process takes place near the $\bar{K} N N$ threshold.  Therefore, the
spinor factor $\left ( \chi _{p}^{\dagger} \chi _{\uparrow} \right )
\left ( \chi _{\Lambda}^{\dagger} \bm{\sigma} \chi _{\downarrow}
\right )$ is the same as in Eq.~\eqref{eq:mathT1}.  Second, the
multiple scattering amplitude of Fig.~\ref{fig:FCA} is calculated
employing the so-called fixed center approximation to the Faddeev
equation~\cite{Bayar:2011qj, Bayar:2012hn}, and we denote this part as
$T^{\rm FCA}$.  Third, the kaon absorption takes place after the
multiple scattering, which is be evaluated in the same manner as the
$K^{-} p \Lambda$ vertex in Eq.~\eqref{eq:mathT1}.  Here, we have two
types of vertices, $K^{-} p \Lambda$ and $\bar{K}^{0} n \Lambda$, and
both have the same structure and coupling constant as in
Eq.~\eqref{eq:Vmbb}.  As a consequence, the scattering amplitude
$\mathcal{T}_{1}$ can be obtained by replacing $T_{2}^{(K^{-} p \to
  K^{-} p)}$ with $T^{\rm FCA}$ in Eq.~\eqref{eq:mathT1} as
\begin{align}
  \mathcal{T}_{1} = & i \left ( \chi _{n}^{\dagger} \chi \right )
  \left ( \chi _{p}^{\dagger} \chi _{\uparrow} \right ) 
  T_{1}^{(K^{-} n \to K^{-} n)} ( w_{1}, \, \cos \theta _{1} )
  \tilde{V} 
  \int \frac{d^{3} q}{( 2 \pi )^{3}} 
  \mathcal{F} ( \bm{q} ) \bm{q}  \left ( \chi _{\Lambda}^{\dagger} \bm{\sigma}
  \chi _{\downarrow} \right )
  \notag \\ & \times
  \left [ 
    \frac{T_{11}^{\rm FCA} + T_{41}^{\rm FCA}}
         {(q^{0})^{2} - \omega _{K^{-}} ( \bm{q} )^{2}}
    + \frac{T_{13}^{\rm FCA} + T_{43}^{\rm FCA}}
    {(q^{0})^{2} - \omega _{\bar{K}^{0}} ( \bm{q} )^{2}}
    \right ]
  \int \frac{d^{3} p}{( 2 \pi )^{3}}
  \frac{\tilde{\Psi} ( p_{\lambda} , \, p_{\rho} ) }
       {(p^{0})^{2} - \omega _{K^{-}} ( \bm{p} )^{2} + i m_{K^{-}} \Gamma _{K}} ,
  \label{eq:mathT1FCA}
\end{align}
where everything, except for the amplitude of the multiple scattering
$T^{\rm FCA}$, is evaluated as in Eq.~\eqref{eq:mathT1}.  In
particular, the energies of the kaons, $p^{0}$ and $q^{0}$, are fixed
by the option A or B.

The multiple scattering amplitude $T_{i j}^{\rm FCA}$ is labeled by
channel indices $i$ and $j$.  Since we have the $\bar{K} N N$ system
decaying into $\Lambda p$, we consider six channels in the order
$K^{-} p p$, $\bar{K}^{0} n p$, $\bar{K}^{0} p n$, $p p K^{-}$, $n p
\bar{K}^{0}$, and $p n \bar{K}^{0}$.  In this study, we neglect
diagrams with pion exchange between two nucleons in the $T_{i j}^{\rm
  FCA}$ amplitude, since these contributions are found to be small
(see Appendix~\ref{app:4}).  We note that we distinguish the ordering
of the kaon and the two nucleons; in channels where the kaon appears
first (last), the kaon interacts with the first (second) nucleon of
the ordering at the first or last of the scatterings.  In this sense,
the $T_{11}^{\rm FCA}$ ($T_{41}^{\rm FCA}$) in
Eq.~\eqref{eq:mathT1FCA} represents the multiple $K^{-} p p$ ($p p
K^{-}$) $\to K^{-} p p$ scattering, which means that the $K^{-}$ first
scatters with the left (right) of two nucleons and last with the left.
In other words, $T_{11}^{\rm FCA}$ and $T_{41}^{\rm FCA}$ comes from
odd- and even-numbered interaction diagrams in the right-hand side of
Fig.~\ref{fig:FCA}.  Then, the sum $T_{11}^{\rm FCA} + T_{41}^{\rm
  FCA}$ represents the whole of the multiple scattering amplitude in
Fig.~\ref{fig:FCA}.  For these terms, the $K^{-}$ is absorbed via the
$K^{-} p \Lambda$ vertex.  Similarly, the term $T_{13}^{\rm FCA}$ and
$T_{43}^{\rm FCA}$ represent $K^{-} p p$ and $p p K^{-} \to
\bar{K}^{0} p n$ scatterings, respectively, and the $\bar{K}^{0}$ is
absorbed via the $\bar{K}^{0} n \Lambda$ vertex.  Since we fix the
ordering of the final state in the reaction as $n p \Lambda$ from left
to right, the final-state channel of the multiple scattering should be
$1$ or $3$, where the kaon is absorbed by the second nucleon of the
pair after the interaction with the first one which emits the proton.
We also note that the multiple scattering amplitude in the fixed
center approximation is a function of the invariant mass $M_{\Lambda
  p}$.  The formulation of the fixed center approximation is given in
Appendix~\ref{app:4}, and the details are given in
Refs.~\cite{Bayar:2011qj, Bayar:2012hn}.

In a similar manner, we can calculate the amplitudes $\mathcal{T}_{2,
  3, 4, 5, 6}$ as:
\begin{align}
  \mathcal{T}_{2} = & - i \left ( \chi _{n}^{\dagger} \chi \right )
  \left ( \chi _{p}^{\dagger} \chi _{\downarrow} \right ) 
  T_{1}^{(K^{-} n \to K^{-} n)} ( w_{1}, \, \cos \theta _{1} )
  \tilde{V} 
  \int \frac{d^{3} q}{( 2 \pi )^{3}}
  \mathcal{F} ( \bm{q} ) \bm{q}  \left ( \chi _{\Lambda}^{\dagger}
  \bm{\sigma} \chi _{\uparrow} \right )
  \notag \\ & \times
  \left [
    \frac{T_{11}^{\rm FCA} + T_{41}^{\rm FCA}}
         {(q^{0})^{2} - \omega _{K^{-}} ( \bm{q} )^{2}} 
    + \frac{T_{13}^{\rm FCA} + T_{43}^{\rm FCA}}
         {(q^{0})^{2} - \omega _{\bar{K}^{0}} ( \bm{q} )^{2}} 
  \right ]
  \int \frac{d^{3} p}{( 2 \pi )^{3}}
  \frac{\tilde{\Psi} ( p_{\lambda} , \, p_{\rho} )}
       {(p^{0})^{2} - \omega _{K^{-}} ( \bm{p} )^{2} + i
         m_{K^{-}} \Gamma _{K}} ,
\end{align}
\begin{align}
  \mathcal{T}_{3} = & - i \left ( \chi _{n}^{\dagger} \chi _{\uparrow} \right )
  \left ( \chi _{p}^{\dagger} \chi \right ) 
  T_{1}^{(K^{-} p \to \bar{K}^{0} n)} ( w_{1}^{\prime}, \, \cos \theta _{1}^{\prime} )
  \tilde{V} 
  \int \frac{d^{3} q}{( 2 \pi )^{3}} 
  \mathcal{F} ( \bm{q} ) \bm{q}  \left ( \chi _{\Lambda}^{\dagger} \bm{\sigma}
  \chi _{\downarrow} \right )
  \notag \\ & \times
  \left [
    \frac{T_{21}^{\rm FCA} + T_{51}^{\rm FCA}}
         {(q^{0})^{2} - \omega _{K^{-}} ( \bm{q} )^{2}}
    + \frac{T_{23}^{\rm FCA} + T_{53}^{\rm FCA}}
         {(q^{0})^{2} - \omega _{\bar{K}^{0}} ( \bm{q} )^{2}}
  \right ]
  \int \frac{d^{3} p}{( 2 \pi )^{3}}
  \frac{\tilde{\Psi} ( p_{\lambda} , \, p_{\rho} )}
       {(p^{\prime \, 0})^{2} - \omega _{\bar{K}^{0}} ( \bm{p} )^{2} + i m_{\bar{K}^{0}}
         \Gamma _{K}}
   ,
\end{align}
\begin{align}
  \mathcal{T}_{4} = & i \left ( \chi _{n}^{\dagger} \chi _{\downarrow} \right )
  \left ( \chi _{p}^{\dagger} \chi \right ) 
  T_{1}^{(K^{-} p \to \bar{K}^{0} n)} ( w_{1}^{\prime}, \, \cos \theta _{1}^{\prime} )
  \tilde{V} 
  \int \frac{d^{3} q}{( 2 \pi )^{3}} 
  \mathcal{F} ( \bm{q} ) \bm{q}  \left ( \chi _{\Lambda}^{\dagger} \bm{\sigma}
  \chi _{\uparrow} \right )
  \notag \\ & \times
  \left [
    \frac{T_{21}^{\rm FCA} + T_{51}^{\rm FCA}}
         {(q^{0})^{2} - \omega _{K^{-}} ( \bm{q} )^{2}}
    + \frac{T_{23}^{\rm FCA} + T_{53}^{\rm FCA}}
         {(q^{0})^{2} - \omega _{\bar{K}^{0}} ( \bm{q} )^{2}}
  \right ]
  \int \frac{d^{3} p}{( 2 \pi )^{3}}
  \frac{\tilde{\Psi} ( p_{\lambda} , \, p_{\rho} )}
       {(p^{\prime \, 0})^{2} - \omega _{\bar{K}^{0}} ( \bm{p} )^{2} + i m_{\bar{K}^{0}}
         \Gamma _{K}}
       ,
\end{align}
\begin{align}
  \mathcal{T}_{5} = & i \left ( \chi _{n}^{\dagger} \chi _{\uparrow} \right )
  \left ( \chi _{p}^{\dagger} \chi _{\downarrow} \right ) 
  T_{1}^{(K^{-} p \to \bar{K}^{0} n)} ( w_{1}^{\prime}, \, \cos \theta _{1}^{\prime} )
  \tilde{V} 
  \int \frac{d^{3} q}{( 2 \pi )^{3}} 
  \mathcal{F} ( \bm{q} ) \bm{q}  \left ( \chi _{\Lambda}^{\dagger} \bm{\sigma}
    \chi \right )
  \notag \\ & \times
  \left [
    \frac{T_{31}^{\rm FCA} + T_{61}^{\rm FCA}}
         {(q^{0})^{2} - \omega _{K^{-}} ( \bm{q} )^{2}}
    + \frac{T_{33}^{\rm FCA} + T_{63}^{\rm FCA}}
         {(q^{0})^{2} - \omega _{\bar{K}^{0}} ( \bm{q} )^{2}}
  \right ]
  \int \frac{d^{3} p}{( 2 \pi )^{3}}
  \frac{\tilde{\Psi} ( p_{\lambda} , \, p_{\rho} )}
       {(p^{\prime \, 0})^{2} - \omega _{\bar{K}^{0}} ( \bm{p} )^{2} + i m_{\bar{K}^{0}}
         \Gamma _{K}}
   ,
\end{align}
\begin{align}
  \mathcal{T}_{6} = & - i \left ( \chi _{n}^{\dagger} \chi _{\downarrow} \right )
  \left ( \chi _{p}^{\dagger} \chi _{\uparrow} \right ) 
  T_{1}^{(K^{-} p \to \bar{K}^{0} n)} ( w_{1}^{\prime}, \, \cos \theta _{1}^{\prime} )
  \tilde{V} 
  \int \frac{d^{3} q}{( 2 \pi )^{3}} 
  \mathcal{F} ( \bm{q} ) \bm{q}  \left ( \chi _{\Lambda}^{\dagger} \bm{\sigma}
    \chi \right )
  \notag \\ & \times
  \left [
    \frac{T_{31}^{\rm FCA} + T_{61}^{\rm FCA}}
         {(q^{0})^{2} - \omega _{K^{-}} ( \bm{q} )^{2}}
    + \frac{T_{33}^{\rm FCA} + T_{63}^{\rm FCA}}
         {(q^{0})^{2} - \omega _{\bar{K}^{0}} ( \bm{q} )^{2}}
  \right ]
  \int \frac{d^{3} p}{( 2 \pi )^{3}}
  \frac{\tilde{\Psi} ( p_{\lambda} , \, p_{\rho} )}
       {(p^{\prime \, 0})^{2} - \omega _{\bar{K}^{0}} ( \bm{p} )^{2} + i m_{\bar{K}^{0}}
         \Gamma _{K}}
       .
\end{align}

\begin{figure}[t]
  \centering
  \Psfig{8.6cm}{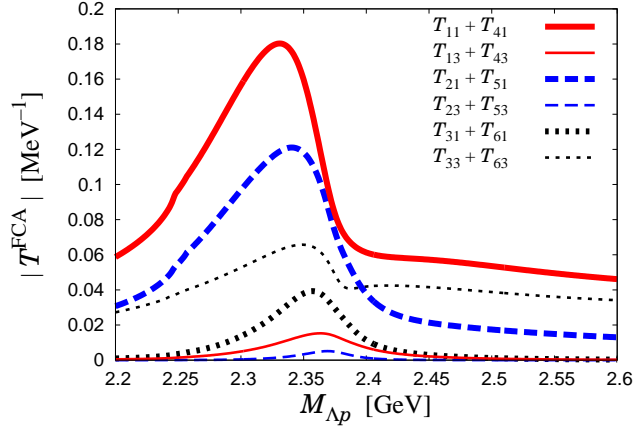}
  \caption{Absolute values of the kaon multiple amplitudes evaluated
    in the fixed center approximation.}
  \label{fig:T_FCA}
\end{figure}

In Fig.~\ref{fig:T_FCA} we show the absolute values of the kaon
multiple amplitudes $T_{11}^{\rm FCA} + T_{41}^{\rm FCA}$,
$T_{13}^{\rm FCA} + T_{43}^{\rm FCA}$, $T_{21}^{\rm FCA} + T_{51}^{\rm
  FCA}$, $T_{23}^{\rm FCA} + T_{53}^{\rm FCA}$, $T_{31}^{\rm FCA} +
T_{61}^{\rm FCA}$, and $T_{33}^{\rm FCA} + T_{63}^{\rm FCA}$, which
are evaluated in the fixed center approximation as functions of
$M_{\Lambda p}$.  As one can see from the figure, the amplitudes
$T^{\rm FCA}$ show a peak around $M_{\Lambda p} = 2.35 \gev$, which
corresponds to a signal of the $\bar{K} N N$ quasi-bound state in our
approach.  Therefore, in the $\Lambda p$ invariant mass spectrum of
the $\HeT ( K^{-} , \, \Lambda p ) n$ reaction, we will observe such a
peak structure if the strength of the peak for the signal of the
$\bar{K} N N$ quasi-bound state is strong enough.  The amplitude $|
T_{11}^{\rm FCA} + T_{41}^{\rm FCA} |$ has the strongest signal peak,
since the $\bar{K} N N$ quasi-bound state is generated dominantly by
the $K^{-} p \to K^{-} p$ scattering in the $K^{-} p p$ configuration.
Here we note that the kaon absorption width $\Gamma_{K}$ is fixed as
15 MeV, so that the amplitude $T_{11}^{FCA} + T_{41}^{FCA}$ reproduces
the width of the $\bar{K} N N$ bound-state signal
in~\cite{Bayar:2012hn}.  We also note that we cannot see a clear
$\bar{K} N N$ threshold around $2.37 \gev$ in the amplitude $T^{\rm
  FCA}$ in Fig.~\ref{fig:T_FCA} because we introduce a finite width
for the kaon propagators of the amplitude.  In contrast, the cusp at
$2.25 \gev$ is caused by the $\pi \Sigma N$ threshold, where the $\pi
\Sigma$ degrees of freedom are intrinsically contained in the two-body
chiral $\bar{K} N \to \bar{K} N$ amplitudes employed in the
construction of the $T^{\rm FCA}$ amplitudes.

\section{Numerical results}
\label{sec:3}

In this section we show the numerical results of our calculations for
the cross section of the $\HeT ( K^{-} , \, \Lambda p ) n$ reaction.
We fix the initial kaon momentum in the laboratory frame to be $k_{\rm
  lab} = 1 \gev /c$, as in the experiment~\cite{Sada:2016vkt}. We note
that the momentum of the intermediate kaon after the first $K^{-} n
\to K^{-} n_{\rm escape}$ or $K^{-} p \to \bar{K}^{0} n_{\rm escape}$
collision becomes $\lesssim 50 \mev /c$ for an initial kaon momentum
of $k_{\rm lab} = 1 \gev /c$, if the escaping neutron goes forward.
This means that the initial $K^{-}$ energy chosen favors the
production of low energy kaons and is suitable to see a possible
$\bar{K} N N$ state around its threshold.

As mentioned before, we consider two scenarios to reproduce a peak
structure in the $\Lambda p$ invariant mass spectrum of the $\HeT (
K^{-} , \, \Lambda p ) n$ reaction.  One consists of generating the
$\Lambda (1405)$ but not allowing it to form a bound state with the
remaining proton. The uncorrelated $\Lambda (1405) p$ system
eventually decays into a $\Lambda p$ pair.  The other consists of
generating a $\bar{K} N N$ quasi-bound state which subsequently decays
into $\Lambda p$.  The numerical results of the spectrum in these two
scenarios are shown in Sec.~\ref{sec:3-1} and Sec.~\ref{sec:3-2},
respectively.

Our numerical results are compared with the experimental
data~\cite{Sada:2016vkt} and in particular to their analysis in terms
of a Breit-Wigner amplitude
\begin{equation}
  \frac{d \sigma}{d M_{\Lambda p}} \propto p_{n}^{\prime} p_{\Lambda}^{\ast}
  \frac{\Gamma _{X}^{2}}{(M_{\Lambda p} - M_{X})^{2} + \Gamma _{X}^{2} / 4} ,
  \label{eq:E15fit}
\end{equation}
with parameters $M_{X} = 2355 \, ^{+6}_{-4} \text{(stat.)} \pm 12
\text{(sys.)}$ MeV and $\Gamma _{X} = 110 \, ^{+19}_{-17}
\text{(stat.)}  \pm 27 \text{(sys.)}$
MeV~\cite{Sada:2016vkt}.\footnote{We note that the parameters for this
  mass spectrum as well as the experimental data shown in the figures
  were obtained without acceptance corrections.} The comparison
between our results and this mass spectrum helps us to discuss the
origin of the peak structure in the invariant mass spectrum.

\subsection{Generating an uncorrelated $\Lambda (1405) p$ system}
\label{sec:3-1}

\begin{figure}[!t]
  \centering
  \Psfig{15.50cm}{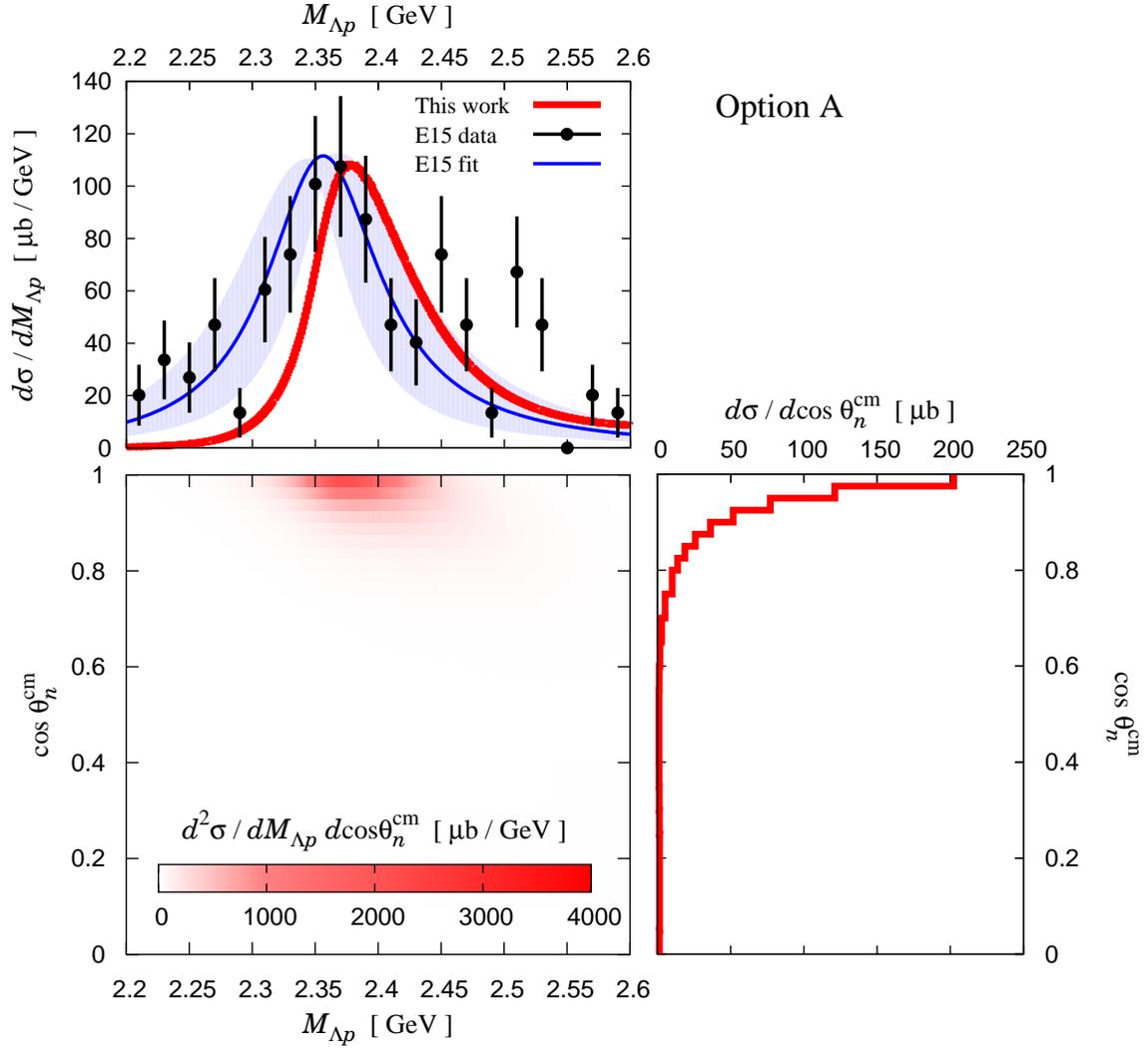} 
  \caption{Differential cross section of the in-flight $\HeT ( K^{-} ,
    \, \Lambda p ) n$ reaction for the uncorrelated $\Lambda (1405) p$
    system (see Fig.~\ref{fig:3NA}) with the option A.  The
    experimental (E15) data for the mass spectrum $d \sigma / d
    M_{\Lambda p}$ and its fit~[Eq.~\eqref{eq:E15fit}] are taken from
    Ref.~\cite{Sada:2016vkt} and are shown in arbitrary units.}
  \label{fig:result_unc_A}
\end{figure}

\begin{figure}[!t]
  \centering
  \Psfig{15.50cm}{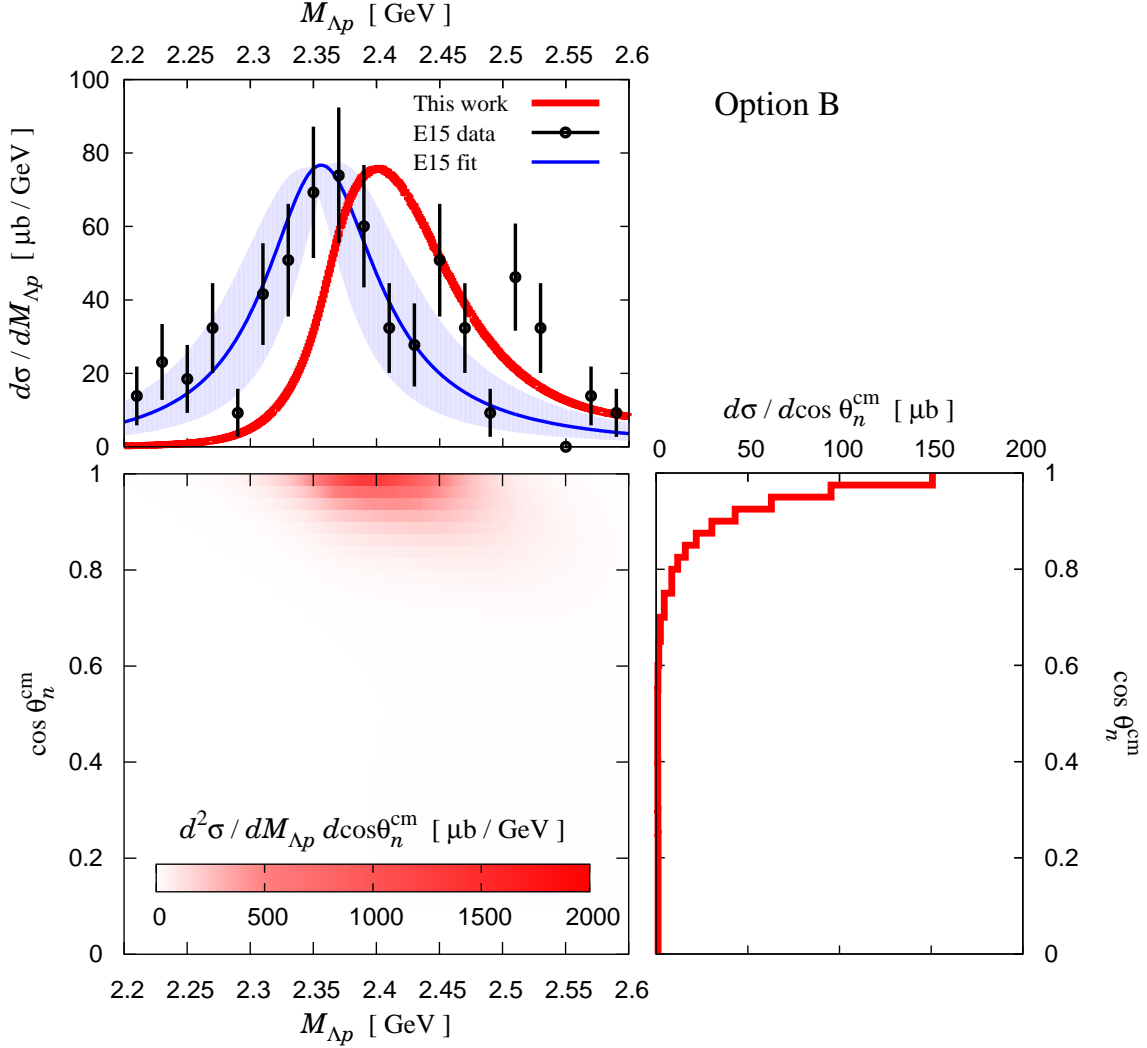}
  \caption{The same as Fig.~\ref{fig:result_unc_A} for the kinematic
    option B. }
  \label{fig:result_unc_B}
\end{figure}

First, we consider the case that the uncorrelated $\Lambda (1405) p$
system is generated in the in-flight $\HeT ( K^{-} , \, \Lambda p ) n$
reaction, as formulated in Sec.~\ref{sec:2-2}.  The numerical
results of the differential cross section are shown in
Figs.~\ref{fig:result_unc_A} and \ref{fig:result_unc_B} for the options
A and B, respectively.

As seen from Figs.~\ref{fig:result_unc_A} and \ref{fig:result_unc_B},
both in the options A and B, we find the peak of the invariant mass
spectrum $d \sigma / d M_{\Lambda p}$ at around the $K^{-} p p$
threshold, $M_{\Lambda p} \approx m_{K^{-}} + m_{p} + m_{p} = 2.370
\gev$.  This corresponds to the signal of the uncorrelated $\Lambda
(1405) p$.  Here we emphasize that the uncorrelated $\Lambda (1405) p$
gives a peak not at the $\Lambda (1405) p$ threshold, which is about
$2.355 \gev$ in our model, but just at the $K^{-} p p$ threshold in
the option A or at $2.4 \gev$ in the option B.  This is because the
$\Lambda (1405)$ is slowly moving due to the momentum $\bm{p}$ carried
by the kaon after the first scattering, and hence the moving $\Lambda
(1405)$ and the proton (the third nucleon of $\HeT$) have an
invariant mass larger than the $\Lambda (1405) p$ threshold.

However, regardless of the option A or B, the peak position is not
consistent with the experimental data and their
fit~[Eq.~\eqref{eq:E15fit}], which are shown in
Figs.~\ref{fig:result_unc_A} and \ref{fig:result_unc_B} in arbitrary
units.  The peak position in the experiment is more than 20 MeV lower
than that of our result in the uncorrelated $\Lambda (1405) p$ case.
In addition, we cannot reproduce the behavior of the tail of the peak
below the $K^{-} p p$ threshold, say at $M_{\Lambda p} = 2.3 \gev$.
This fact indicates that the experimental signal observed in the
in-flight $\HeT ( K^{-} , \, \Lambda p ) n$ reaction at
J-PARC~\cite{Sada:2016vkt} is definitely not the uncorrelated $\Lambda
(1405) p$ state.

In Figs.~\ref{fig:result_unc_A} and \ref{fig:result_unc_B}, we also
show the behavior of the angular distribution of the cross section $d
\sigma / d \cos \theta _{n}^{\rm cm}$ as well as the double
differential cross section $d^{2} \sigma / d M_{\Lambda p} d \cos
\theta _{n}^{\rm cm}$.  Here we show the results only in the region
$\cos \theta _{n}^{\rm cm} \ge 0$, since there is no significant
contribution in the region $\cos \theta _{n}^{\rm cm} < 0$.  From the
figure, one can see that the uncorrelated $\Lambda (1405) p$ signal
dominantly comes from the condition of forward neutron emission, i.e.
$\theta _{n}^{\rm cm} \approx 0$ degrees.  This is caused by both
kinematic and dynamical reasons.  As far as kinematics is concerned,
we note that for finite scattering angle, $\theta _{n}^{\rm cm} > 0$
degree, the kaon momenta in the intermediate states $| \bm{p} |$ and
$| \bm{q} |$ become large and are comparable to the initial kaon
momentum $1 \gev /c$ in the laboratory frame.  Therefore, the kaon
propagator $1 / [ (q^{0})^{2} - \bm{q}^{2} - m_{K}^{2} ]$ and the form
factor $\mathcal{F} (\bm{q})$ suppress the cross section for $\theta
_{n}^{\rm cm} > 0$ degree\footnote{Since $(q^{0})^{2} - m_{K}^{2}$ is
  always negative in the present kinematics, the square of this
  propagator, $1 / [ (q^{0})^{2} - \bm{q}^{2} - m_{K}^{2} ]^{2}$,
  monotonically decreases as $| \bm{q} |$ increases.}.  With respect
to the dynamics, we observe that the cross sections of $K^{-} p \to
\bar{K}^{0} n$ and $K^{-} n \to K^{-} n$ at $k_{\rm lab} = 1 \gev /c$
have a local maximum for forward neutron emission, as seen in
Fig.~\ref{figA:dsdO} in Appendix~\ref{app:2}, and hence the forward
neutron emission is also favored in the $\HeT ( K^{-} , \, \Lambda p )
n$ reaction.  Because of these reasons, the uncorrelated $\Lambda
(1405) p$ signal favors the condition of forward neutron emission,
$\theta _{n}^{\rm cm} \approx 0$ degrees\footnote{This means that the
  neutron goes in opposite direction to the original one in the $K^{-}
  n \to K^{-} n$ center-of-mass frame, and equivalently the kaon goes
  backward in that frame.}.  We also note that the peak shifts
slightly upwards as the scattering angle $\theta _{n}^{\rm cm}$
increases, which means that the $\Lambda (1405)$ gets more momentum
from the kaon after the first scattering.  This contribution can be
seen as the band from the $\bar{K} N N$ threshold at $\cos \theta
_{n}^{\rm cm} = 1$ to the lower-right direction in the $d^{2} \sigma /
d M_{\Lambda p} d \cos \theta _{n}^{\rm cm}$ plot of the figures,
although its strength is very weak due to the kinematic and dynamical
reasons explained above.

As for the dependence of the cross section on the cutoff of the
form-factor employed in the $K^{-} p \Lambda$
vertex~\eqref{eq:FF_vertex}, we have found that $\Lambda$ values in
the range $[0.7 \gev , \, 1.0 \gev ]$ produce only quantitative
differences in the corresponding mass spectra and angular
distributions, while their shape is preserved.  In particular, with
the cutoff value $\Lambda = 1.0 \gev$ the height of the mass spectrum
becomes about $1.3$ times larger than that with $\Lambda = 0.8 \gev$
in both options A and B, and likewise the size of the spectra gets
reduced by about $20 \%$ for a cutoff value of $\Lambda = 0.7 \gev$.

Let us discuss the difference between the results in the options A and
B.  In the option A (B), the peak height of the mass spectrum $d
\sigma / d M_{\Lambda p}$ is about $110 \microb / \text{GeV}$ ($80
\microb / \text{GeV}$), and the peak position is about $M_{\Lambda p}
= 2.37 \gev$ ($2.4 \gev$).  This difference of the peak structure is
caused by the treatment of the kaon energies in the intermediate state
and could be interpreted as a theoretical ambiguity in calculating the
mass spectrum of this reaction in the present formulation.

\begin{figure}[!t]
  \centering
  \Psfig{8.6cm}{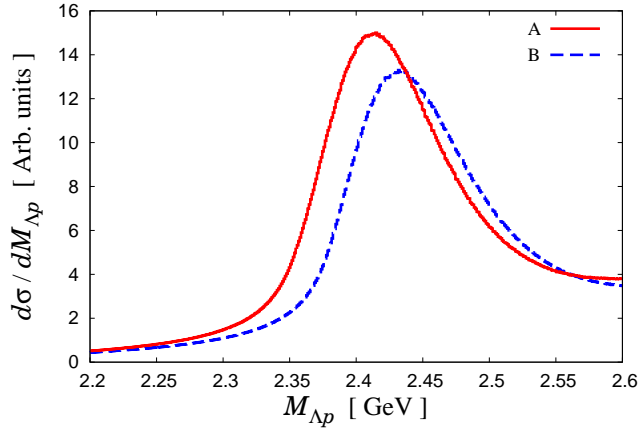}
  \caption{Mass spectrum for the $\Lambda p$ invariant mass of the
    in-flight $\HeT ( K^{-} , \, \Lambda p ) n$ reaction with a
    constant $T_{2}$.}
  \label{fig:dsdM_base}
\end{figure}

In order to see the structure created by the underlying kinematic
features of the amplitudes rather than by the uncorrelated $\Lambda
(1405) p$ system, we take the amplitude $T_{2}$ as a constant:
\begin{equation}
  T_{2}^{(K^{-} p \to K^{-} p)} = T_{2}^{(\bar{K}^{0} n \to K^{-} p)}
  = \text{const}.
\end{equation}
The result of the mass spectrum with this constant amplitude is
plotted in Fig.~\ref{fig:dsdM_base}.  As one can see, even if we do
not take into account the $\Lambda (1405)$ contribution, a peak
appears in the $M_{\Lambda p}$ mass spectrum just above $2.4 \gev$.
This is due to the quasi-elastic scattering of the kaon in the first
collision of the process.  Namely, the intermediate kaon after the
neutron emission at $T_{1}$ goes almost to its on mass shell, where
the denominator of the propagator $1/[(p^{(\prime ) 0})^{2} -
\bm{p}^{2} - m_{K}^{2} + i m_{K} \Gamma _{K}]$ gets close to zero.
Then, the peak position and its height is slightly different in the
options A and B.  Since the energy $p^{( \prime ) 0}$ contains the
kinetic energies of the nucleons with negative signs in the option B,
the denominator, $(p^{(\prime ) 0})^{2} - \bm{p}^{2} - m_{K}^{2} + i
m_{K} \Gamma _{K}$, gets close to zero for a larger value of
$p_{\Lambda}^{\prime \, 0} + p_{p}^{\prime \, 0}$ in this option.
This kinematic fact makes the peak in option B to appear at a higher
energy compared to that in A.  We emphasize that this shift of the
peak position makes a significant difference in the signal region.
Actually, at the $\bar{K} N N$ threshold the mass spectrum in
Fig.~\ref{fig:dsdM_base} can be twice as large in option A than in
option B.  As a consequence, the uncorrelated $\Lambda (1405) p$
contribution, which can be calculated essentially by the product of
the squared amplitude $| T_{2} |^{2}$ and the mass spectrum in
Fig.~\ref{fig:dsdM_base} and whose peak position eventually appears
around the $\bar{K} N N$ threshold, shows a non-negligible difference
in both options.

The other lesson that we learn from this exercise is that the peaks in
Fig.~\ref{fig:dsdM_base} are shifted upwards by about 30 MeV with
respect to those in Figs.~\ref{fig:result_unc_A} and
\ref{fig:result_unc_B}.  This reflects the fact that the excitation of
the $\Lambda (1405)$ in Figs.~\ref{fig:result_unc_A} and
\ref{fig:result_unc_B} puts strength to the left of the quasi-elastic
kaon peak, which makes it appear at lower energies merged with the
signal of the uncorrelated $\Lambda (1405)p$ pair.

\subsection{Generating a $\bar{K} N N$ quasi-bound state}
\label{sec:3-2}

\begin{figure}[!t]
  \centering
  \Psfig{15.50cm}{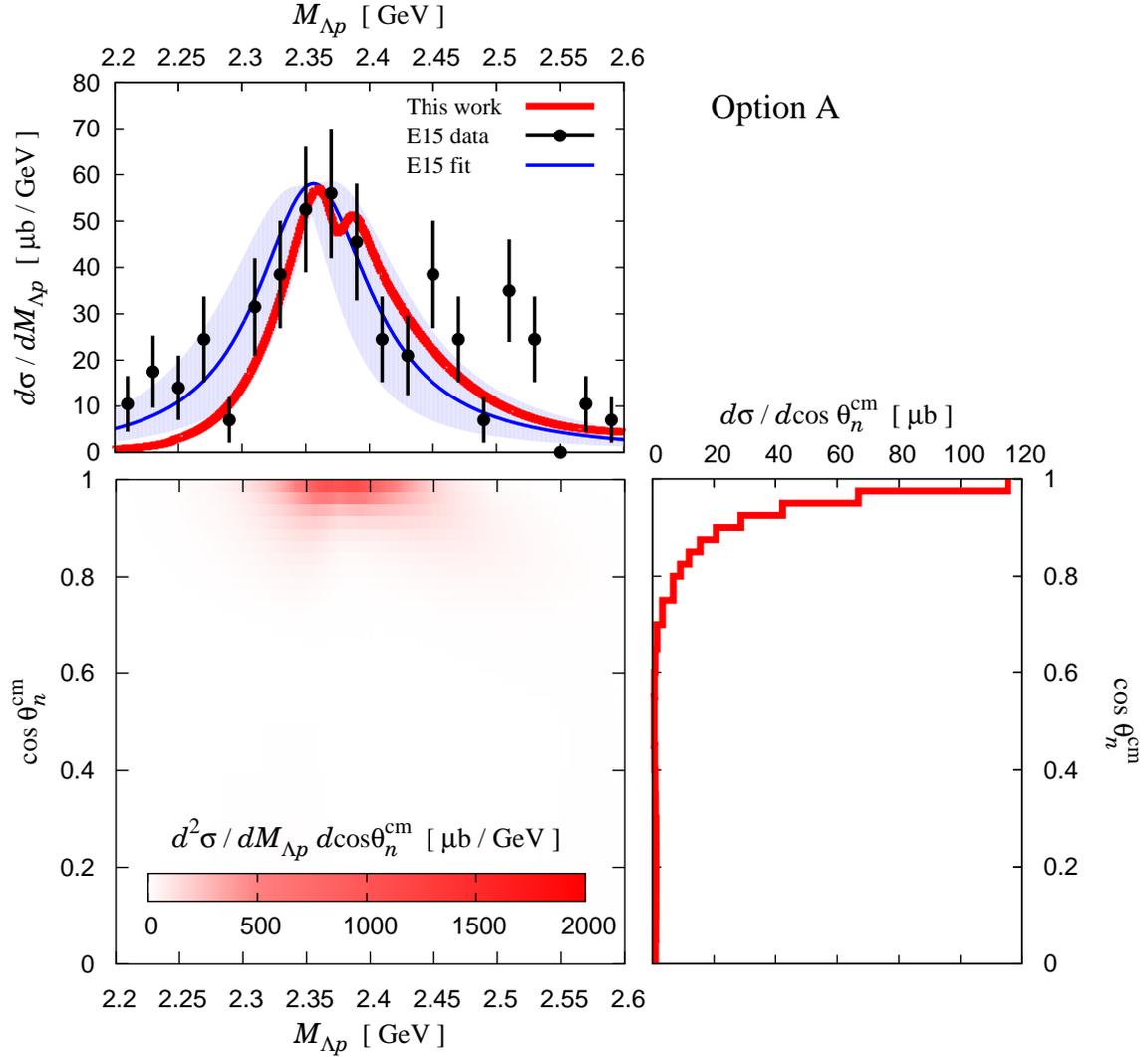} 
  \caption{Differential cross section for the in-flight $\HeT ( K^{-}
    , \, \Lambda p ) n$ reaction with generating the $\bar{K} N N$
    quasi-bound state (see Fig.~\ref{fig:3NA_FCA}) in the option A.
    The experimental (E15) data and its fit in the mass spectrum $d
    \sigma / d M_{\Lambda p}$ are taken from Ref.~\cite{Sada:2016vkt}
    and are shown in arbitrary units.}
  \label{fig:result_KbarNN_A}
\end{figure}

\begin{figure}[!t]
  \centering
  \Psfig{15.50cm}{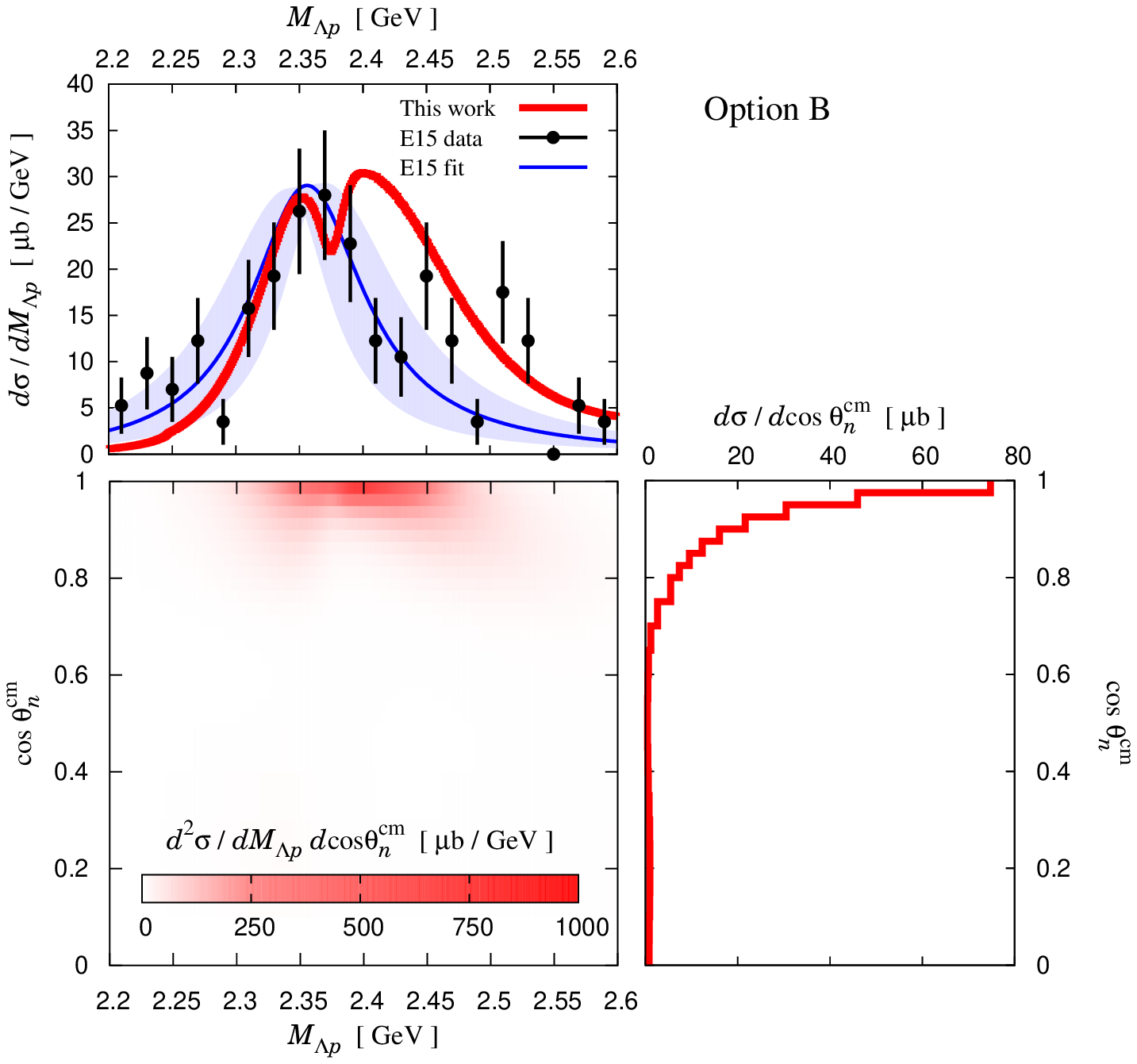}
  \caption{The same as Fig.~\ref{fig:result_KbarNN_A} for the
    kinematic option B.}
  \label{fig:result_KbarNN_B}
\end{figure}

In the previous subsection we have shown that the experimental signal
observed in the in-flight $\HeT ( K^{-} , \, \Lambda p ) n$ reaction
at J-PARC~\cite{Sada:2016vkt} is definitely not the uncorrelated
$\Lambda (1405) p$ state.  Here we consider the case that a $\bar{K} N
N$ quasi-bound state is generated after the first kaon scattering
$T_{1}$, as formulated in Sec.~\ref{sec:2-3}.  The numerical results
are shown in Figs.~\ref{fig:result_KbarNN_A} and
\ref{fig:result_KbarNN_B}.

An important thing to be noted is that the peak structure in our mass
spectrum, regardless of the option A or B, is consistent
with the experimental results.  In particular, we can reproduce
qualitatively well the behavior of the tail of the peak below the
$K^{-} p p$ threshold, say  at $M_{\Lambda p} = 2.3 \gev$.  We also note
that the width of our peak is similar to that in the experiment.
Therefore, our mass spectrum supports the explanation that the peak structure observed
in the experiment of Ref.~\cite{Sada:2016vkt} is indeed a signal of the $\bar{K} N
N$ quasi-bound state.

Besides, interestingly, we observe a two-peak structure of the mass
spectrum around the $\bar{K} N N$ threshold from the figures in both
options A and B.  The peak just below the $\bar{K} N N$ threshold
corresponds to the $\bar{K} N N$ quasi-bound state, while the second
peak just above the $\bar{K} N N$ threshold comes from the
quasi-elastic scattering of the kaon, as seen in
Fig.~\ref{fig:dsdM_base}.  Note that, since the mass spectrum is
essentially a product of the mass spectrum in Fig.~\ref{fig:dsdM_base}
and the square of the multiple scattering amplitude, $| T^{\rm FCA}
|^{2}$, the obtained mass spectrum has the peak associated to that of
$| T^{\rm FCA} |^{2}$ and another peak corresponding to that of the
quasi-elastic kaon scattering, which is shifted to slightly lower
energies by the effect of the energy dependence of $| T^{\rm FCA}
|^{2}$ above its peak.

We emphasize that we cannot generate such a two-peak structure when we
consider the case of the uncorrelated $\Lambda (1405) p$ system in the
previous subsection.  In principle, we would always have a possibility
of having such a two-peak structure even in the uncorrelated $\Lambda
(1405) p$ system: one associated with the quasi-elastic scattering of
the kaon, i.e., the first intermediate antikaon practically on-shell,
and another structure linked with the ``$\Lambda (1405)$ signal'' of
$T_{2}$.  However, the location of these structures and their
particular shapes in the uncorrelated $\Lambda (1405) p$ case make
them merge into just one broader peak.

In contrast, for the case of the $\bar{K} N N$ quasi-bound state the
locations of the peaks from the two origins are separated enough such
that a two-peak structure remains.  The fact that the spectrum in
Fig.~\ref{fig:dsdM_base} falls fast to the left, while the dominant
$K^{-} p p$ component ($T_{11}^{\rm FCA} + T_{41}^{\rm FCA}$) in
Fig.~\ref{fig:T_FCA} falls fast to the right, also help in creating a
dip in between the two peaks. This discussion indicates that, if such
a two-peak structure would be observed in experiments, this could be a
strong evidence that there should be a certain state originating from
a dynamical factor, such as a $\bar{K} N N$ bound state, in addition
to the quasi-elastic scattering of the kaon.

Note that the peak height of the mass spectrum in the case of the
$\bar{K} N N$ bound state becomes about half of that of the
uncorrelated $\Lambda (1405) p$ case.  This is due to the combination
of the structure tied to quasi-elastic kaon scattering in the first
collision, shown in Fig.~\ref{fig:dsdM_base}, with that of the
scattering amplitude, which is either $T_{2}$, in the uncorrelated
$\Lambda (1405) p$ case, or $T^{\rm FCA}$, in the $\bar{K} N N$ case.
Hence, the peak produced by the $T_2$ or $T^{\rm FCA}$ amplitude will
be multiplied by the quasi-elastic kaon scattering structure of
Fig.~\ref{fig:dsdM_base}.  Actually, although the peak heights of
$T_{2}$ and $T^{\rm FCA}$ are similar, their peak positions are
displaced energetically, and this fact produces a drastic effect in
the final value of the corresponding spectrum.  Indeed, the peak of
$T_2$ for the uncorrelated $\Lambda (1405) p$ case appears, due to
some intrinsic momentum of the generated $\Lambda (1405)$, at
invariant $M_{\Lambda p}$ masses above the $\bar{K} N N$ threshold and
close to $2.4 \gev$, which is the position of the kinematical peak
associated to quasi-elastic scattering, hence producing an enhanced
effect in the resulting mass spectrum.  This is opposite in the
correlated case producing the $\bar{K} N N$ bound state, since the
$T^{\rm FCA}$ amplitude peaks around $2.35 \gev$, a region where the
strength of the kaon quasi-elastic structure has fallen down
appreciably with respect to its peak.  This explains why the mass
spectrum obtained in the case of the $\bar{K} N N$ bound state becomes
half of that for the uncorrelated $\Lambda (1405) p$ case.

In Figs.~\ref{fig:result_KbarNN_A} and \ref{fig:result_KbarNN_B}, we
also plot the angular distribution of the cross section $d \sigma /
d \cos \theta _{n}^{\rm cm}$ as well as the double differential cross
section $d^{2} \sigma / d M_{\Lambda p} d \cos \theta _{n}^{\rm cm}$.
Again there is no significant contribution in the region $\cos \theta
_{n}^{\rm cm} < 0$.  From these results, we can see that the structure
at the $\bar{K} N N$ threshold is generated dominantly in the
condition of forward neutron scattering.  The reason is the same as
that discussed in the previous subsection.  In addition, we may
observe two bands in the $d^{2} \sigma / d M_{\Lambda p} d \cos
\theta _{n}^{\rm cm}$ plot of the figures, although their strength is
weak; one goes from $M_{\Lambda p} \approx 2.35 \gev$ at $\cos \theta
_{n}^{\rm cm} = 1$ to the lower direction, and the other goes from the
$\bar{K} N N$ threshold at $\cos \theta _{n}^{\rm cm} = 1$ to the
lower-right direction.  Actually, the former is the signal of the
$\bar{K} N N$ quasi-bound state, and the latter is the contribution
from the quasi-elastic scattering of the kaon.

Similarly to what we find in the uncorrelated $\Lambda (1405) p$ case,
the use of different cutoff values only affects the size of the mass
spectrum and angular distribution of the $\bar{K} N N$ bound state
signal. More specifically, using a cutoff value of $\Lambda = 1.0
\gev$ ($0.7 \gev$) gives rise to a mass spectrum which is about $1.3$
times higher ($1.2$ times lower) than that with $\Lambda = 0.8 \gev$
in both options A and B.

\begin{figure}[!t]
  \centering \Psfig{8.6cm}{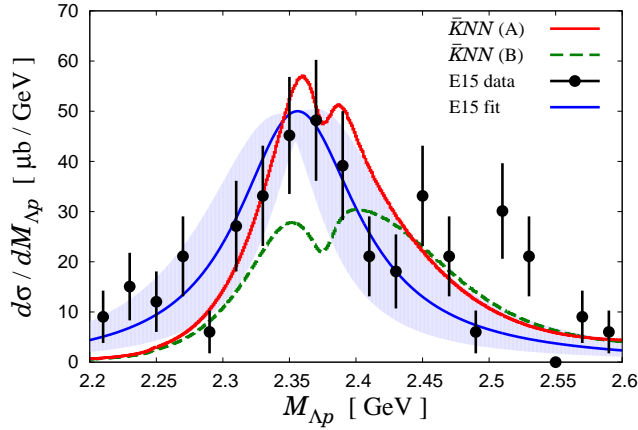}
  \caption{Mass spectrum for the $\Lambda p$ invariant mass of the
    in-flight $\HeT ( K^{-} , \, \Lambda p ) n$ reaction.  The E15 fit
    is scaled so as to reproduce the experimental value $7
    \microb$~\cite{Sada:2016vkt}.}
  \label{fig:dsdM_compare}
\end{figure}

Finally, we compare the results of the mass spectrum employing the
options A and B in Fig.~\ref{fig:dsdM_compare}.  As one can see, the
strength of the peak is different in the two options, although their
distributions become similar at the tails. In fact, the total cross
section, calculated by integrating the mass spectrum $d \sigma / d
M_{\Lambda p}$ with respect to the invariant mass $M_{\Lambda p}$,
gives $7.6 \microb$ for option A and $5.6 \microb$ for option B.  This
difference of the strength can be explained from the results displayed
in Fig.~\ref{fig:dsdM_base}: the different location of the
quasi-elastic kaon scattering peak produces a larger strength for
option A than for option B.

In Fig.~\ref{fig:dsdM_compare} we also plot the J-PARC E15 data and
its fit, which is scaled so as to reproduce the empirical value of $7
\microb$~\cite{Sada:2016vkt}.  We find that the absolute value of our
mass spectrum is qualitatively consistent with the experimental
one.\footnote{Note that one can only aim at a qualitative comparison
  since the J-PARC E15 experiment did not perform acceptance
  corrections so far.}

\section{Conclusion}
\label{sec:4}

We have studied the reaction $\HeT (K^{-}, \Lambda p) n$ measured
recently in the J-PARC E15 experiment for a kaon beam of $1
\gev/c$.  This momentum is suited to produce secondary kaons nearly stopped when
the neutrons go fast and forward in the laboratory system, which is
guaranteed by a form factor that suppresses high momentum kaons in
the intermediate state and $K^{-} n \to K^{-} n$ or $K^{-} p \to
\bar{K}^{0} n$ cross sections that peak at backward
kaon angles in the center of mass frame.

We have concentrated our study on mechanisms which involve the participation of the
 three nucleons in $\HeT$, allowing us to compare our
results with the part of the experimental spectrum where there are no spectator
nucleons.  Our approach relies on a first collision of the
$K^{-}$ with a nucleon in $\HeT$, leading to the production of a slow
$\bar{K}$ that is absorbed by the remaining $N N$ pair.  However,
before the absorption takes place, the kaon is allowed to interact repeatedly with
this pair of nucleons, hence providing a source of binding for the  $\bar{K} N
N$ system.

Technically, we employ a fully antisymmetrized $\HeT$ wave function, which
leads to many different combinations of first and second scattering processes,
and  we use Jacobi coordinates to describe the $\HeT$ system. The
$\bar{K} N$ interaction around threshold is obtained within a chiral unitary
approach, and the interaction of the kaon with the two nucleons is treated in terms
of  the fixed center approximation to the Faddeev equation, which has
proved to be fair enough to deal with this problem.

The results that we obtain are interesting.  We observe that the
$\Lambda p$ invariant mass distribution is clearly shifted to higher
energies compared to experiment when the interaction of the slow kaon
with the remaining two nucleons is not taken into account.  The
situation changes drastically when this interaction is considered,
allowing for the formation of a $\bar{K} N N$ quasi-bound state, and
producing a spectrum which reproduces the shape, position, and width
of the experimental distribution. The total cross section obtained is
also in good agreement with experiment within uncertainties.

A novel aspect of our approach is the presence of a two-peak structure
in the $\Lambda p$ invariant mass spectrum, for which we provide a
justification in terms of a quasi-elastic peak for $\bar{K}$
production in the first collision of the reaction and a peak
associated to the production of the $\bar{K} N N$ bound state that
decays into $\Lambda p$. The present data, which have large error
bars, does not permit to infer this behavior.  It will be interesting
to see what the coming update of the experiment, with larger
statistics, provides.

In any case, from the analysis of this experiment and comparison with
the data, we can claim that the peak seen is indeed tied to the
production of a quasi-bound $\bar{K} N N$ state, with properties very
similar to those obtained in the work of~\cite{Bayar:2012hn} and to the
average of other theoretical works.

\section*{Acknowledgments}

The authors greatly acknowledge Y.~Sada, T.~Yamaga, F.~Sakuma,
J.~Zmeskal, and M.~Iwasaki for discussions on the data of the J-PARC
E15 experiment.
We acknowledge the support by Open Partnership Joint Projects of JSPS
Bilateral Joint Research Projects.
This work is partly supported by Grants-in-Aid for Scientific Research
from MEXT and JSPS (No.~15K17649 and No.~15J06538), the Spanish
Ministerio de Economia y Competitividad 
under the project MDM-2014-0369 of ICCUB and,  with additional FEDER funds,
under 
the contracts FIS2011-28853-C02-01 and FIS2014-54762-P, by the Generalitat
Valenciana in the program Prometeo II, 2014/068, by the Ge\-ne\-ra\-li\-tat de Catalunya contract 2014SGR-401,
and 
by the Spanish Excellence Network on Hadronic Physics FIS2014-57026-REDT.

\appendix

\section{The wave function of ${}^{3}\text{He}$}
\label{app:1}

In this study, we evaluate the wave function of the ${}^{3}\text{He}$
nucleus with the harmonic oscillator potential governed by the
following Hamiltonian:
\begin{equation}
  \hat{H} = \sum _{i = 1}^{3} \frac{\hat{\bm{p}}_{i}^{2}}{2 m_{i}}
  + \sum _{i < j} \frac{k}{2} | \bm{r}_{i} - \bm{r}_{j} |^{2} ,
\end{equation}
where $\hat{\bm{p}}_{i}$ and $\bm{r}_{i}$ are the momentum operator
and coordinate for the $i$th nucleon, respectively, and $k$ is the
spring constant taken as a parameter of the system.  The mass of the nucleon
is fixed as $m_{1} = m_{2} = m_{3} = m_{N}$.

In order to separate the center-of-mass motion, we introduce the
Jacobi coordinates
\begin{equation}
  \begin{split}
    &
    \bm{R} \equiv \frac{m_{1} \bm{r}_{1} + m_{2} \bm{r}_{2} + m_{3} \bm{r}_{3}}
       {m_{1} + m_{2} + m_{3}}
       = \frac{\bm{r}_{1} + \bm{r}_{2} + \bm{r}_{3}}{3} ,
    \\
    &
    \bm{\lambda}
    \equiv \bm{r}_{1} - \frac{m_{2} \bm{r}_{2} + m_{3} \bm{r}_{3}}{m_{2} + m_{3}}
    = \bm{r}_{1} - \frac{\bm{r}_{2} + \bm{r}_{3}}{2} ,
    \quad
    \bm{\rho} \equiv \bm{r}_{3} - \bm{r}_{2} ,
  \end{split}
\end{equation}
and their conjugate momenta
\begin{equation}
  \begin{split}
    & \hat{\bm{P}} \equiv
    \hat{\bm{p}}_{1} + \hat{\bm{p}}_{2} + \hat{\bm{p}}_{3} ,
    \\
    & \hat{\bm{p}}_{\lambda}
    \equiv \frac{(m_{2} + m_{3} )\hat{\bm{p}}_{1}
      - m_{1} ( \hat{\bm{p}}_{2} + \hat{\bm{p}}_{3} )}
           {m_{1} + m_{2} + m_{3}}
    = \frac{2 \hat{\bm{p}}_{1} - \hat{\bm{p}}_{2} - \hat{\bm{p}}_{3}}{3} ,
    \quad
    \hat{\bm{p}}_{\rho}
    \equiv \frac{m_{2} \hat{\bm{p}}_{3} - m_{3} \hat{\bm{p}}_{2}}
           {m_{2} + m_{3}}
    = \frac{\hat{\bm{p}}_{3} - \hat{\bm{p}}_{2}}{2} ,
  \end{split}
\end{equation}
respectively.  With these Jacobi coordinates, we can rewrite the
kinetic part of the Hamiltonian, regardless of values of the masses,
as
\begin{equation}
  \sum _{i = 1}^{3} \frac{\hat{\bm{p}}_{i}^{2}}{2 m_{i}}
  = \frac{\hat{\bm{P}}^{2}}{2 M}
  + \frac{\hat{\bm{p}}_{\lambda}^{2}}{2 m_{\lambda}}
  + \frac{\hat{\bm{p}}_{\rho}^{2}}{2 m_{\rho}}
\end{equation}
where 
\begin{equation}
  \begin{split}
    & M \equiv m_{1} + m_{2} + m_{3} = 3 m_{N} ,
    \\ 
    & m_{\lambda} \equiv \frac{m_{1} ( m_{2} + m_{3} )}{M}
    = \frac{2}{3} m_{N} ,
    \quad 
    m_{\rho} \equiv \frac{m_{2} m_{3}}{m_{2} + m_{3}}
    = \frac{1}{2} m_{N} .
  \end{split}
\end{equation}
On the other hand, in the potential term the $\lambda$ and $\rho$
modes decouple when $m_{2} = m_{3}$, which is the case for
${}^{3}\text{He}$, and its expression is
\begin{equation}
  \sum _{i < j} \frac{k}{2} | \bm{r}_{i} - \bm{r}_{j} |^{2}
  = \frac{1}{2} m_{\lambda} \omega _{\lambda}^{2} \bm{\lambda} ^{2}
  + \frac{1}{2} m_{\rho} \omega _{\rho}^{2} \bm{\rho}^{2} ,
\end{equation}
where the oscillator frequency is defined as
\begin{equation}
  \omega _{\lambda} \equiv 
  \sqrt{\frac{2 k}{m_{\lambda}}} = \sqrt{\frac{3 k}{m_{N}}} ,
  \quad
  \omega _{\rho} \equiv
  \sqrt{\frac{3 k}{2 m_{\rho}}} = \sqrt{\frac{3 k}{m_{N}}} .
\label{eqA:omega}
\end{equation}
As a consequence, the Hamiltonian can be rewritten as
\begin{equation}
  \hat{H} = \frac{\hat{\bm{P}}^{2}}{2 M}
  + \frac{\hat{\bm{p}}_{\lambda}^{2}}{2 m_{\lambda}}
  + \frac{\hat{\bm{p}}_{\rho}^{2}}{2 m_{\rho}}
  + \frac{1}{2} m_{\lambda} \omega _{\lambda}^{2} \bm{\lambda} ^{2}
  + \frac{1}{2} m_{\rho} \omega _{\rho}^{2} \bm{\rho}^{2} .
\end{equation}

Now let us omit the center-of-mass motion and evaluate the wave
function in terms of the $\lambda$ and $\rho$ modes.  For the nucleons in
${}^{3} \text{He}$, we assume that all of them are in the $s$-wave
state and neglect contributions from higher partial waves such as $d$
wave, which are known to be small.  In this condition, the wave
function is expressed as the product of harmonic oscillator wave
functions for  the $\lambda$ and $\rho$ modes, both in the ground state:
\begin{align}
  \Psi ( \lambda , \, \rho ) =
  & 
  \left ( \frac{m_{\lambda} \omega _{\lambda}}{\pi} \right )^{3/4}
  \exp \left ( - \frac{1}{2} m_{\lambda} \omega _{\lambda} \lambda ^{2} \right )
  \times 
  \left ( \frac{m_{\rho} \omega _{\rho}}{\pi} \right )^{3/4}
  \exp \left ( - \frac{1}{2} m_{\rho} \omega _{\rho} \rho ^{2} \right ) ,
\end{align}
with $\lambda \equiv | \bm{\lambda} |$ and $\rho \equiv | \bm{\rho}
|^{2}$.  The normalization of the wave function is
\begin{equation}
  \int d^{3} \lambda  \int d^{3} \rho \, 
  | \Psi ( \lambda , \, \rho ) |^{2} = 1 .
\end{equation}
From the wave function, we can calculate the mean squared radius of
the ${}^{3} \text{He}$ nucleus as the expectation value of $r_{i}^{2}$
for the $i$th nucleon measured from the center of mass.  Namely,
\begin{equation}
  \begin{split}
    & \langle r_{1}^{2} \rangle
    \equiv \int d^{3} \lambda \int d^{3} \rho \,
    | \bm{r}_{1} - \bm{R} | ^{2} | \Psi ( \lambda , \, \rho ) |^{2}
    = \int d^{3} \lambda \int d^{3} \rho 
    \left | \frac{m_{2} + m_{3}}{M} \bm{\lambda} \right | ^{2}
    | \Psi ( \lambda , \, \rho ) |^{2} ,
    \\
    & \langle r_{2}^{2} \rangle
    \equiv \int d^{3} \lambda \int d^{3} \rho \,
    | \bm{r}_{2} - \bm{R} | ^{2} | \Psi ( \lambda , \, \rho ) |^{2}
    = \int d^{3} \lambda \int d^{3} \rho 
    \left | - \frac{m_{1}}{M} \bm{\lambda}
    - \frac{m_{3}}{m_{2} + m_{3}} \bm{\rho} \right | ^{2}
    | \Psi ( \lambda , \, \rho ) |^{2} ,
    \\
    & \langle r_{3}^{2} \rangle
    \equiv \int d^{3} \lambda \int d^{3} \rho \,
    | \bm{r}_{2} - \bm{R} | ^{2} | \Psi ( \lambda , \, \rho ) |^{2}
    = \int d^{3} \lambda \int d^{3} \rho 
    \left | - \frac{m_{1}}{M} \bm{\lambda}
    + \frac{m_{2}}{m_{2} + m_{3}} \bm{\rho} \right | ^{2}
    | \Psi ( \lambda , \, \rho ) |^{2} .
  \end{split}
\end{equation}
A straightforward calculation gives
\begin{equation}
  \langle r_{1}^{2} \rangle
  = \left ( \frac{m_{2} + m_{3}}{M} \right )^{2}
  \frac{3}{2 m_{\lambda} \omega _{\lambda}}
  = \sqrt{\frac{1}{3 k m_{N}}} , 
\end{equation}
\begin{align}
  \langle r_{2}^{2} \rangle = \langle r_{3}^{2} \rangle
  & = \left ( \frac{m_{1}}{M} \right )^{2}
  \frac{3}{2 m_{\lambda} \omega _{\lambda}}
  + \left ( \frac{m_{3}}{m_{2} + m_{3}} \right )^{2}
  \frac{3}{2 m_{\rho} \omega _{\rho}}
  = \sqrt{\frac{1}{3 k m_{N}}} .
\end{align}
As expected, one obtains the same value of the mean squared radius
for each nucleon with $m_{1} = m_{2} = m_{3} = m_{N}$.  Then the
parameter $k$ can be fixed with the empirical value of the mean
squared radius of ${}^{3} \text{He}$, $\langle r_{i}^{2} \rangle =
3.2 \fm ^{2}$.  The result is
\begin{equation}
  k = \frac{1}{3 \langle r_{i}^{2} \rangle ^{2} m_{N}}
  \approx 1.3 \mev / \text{fm} ^{2} .
\end{equation}

In this study we use the ${}^{3} \text{He}$ wave function in momentum
space to calculate the scattering amplitude.  For the harmonic
oscillator potential, we can easily evaluate the wave function in
momentum space as
\begin{align}
  \tilde{\Psi} ( p_{\lambda} , \, p_{\rho} )
  = & \left ( \frac{4 \pi}{m_{\lambda} \omega _{\lambda}} \right ) ^{3/4}
  \exp \left ( - \frac{p_{\lambda}^{2}}
       {2 m_{\lambda} \omega _{\lambda}} \right )
  \times \left ( \frac{4 \pi}{m_{\rho} \omega _{\rho}} \right ) ^{3/4}
  \exp \left ( - \frac{p_{\rho}^{2}}
       {2 m_{\rho} \omega _{\rho}} \right ),
       \label{eqA:tildePsi}
\end{align}
where $p_{\lambda} \equiv | \bm{p}_{\lambda} |$ and $p_{\rho} \equiv |
\bm{p}_{\rho} |$.  This is related to the wave function in coordinate
space as
\begin{equation}
  \tilde{\Psi} ( p_{\lambda} , \, p_{\rho} )
  = \int d^{3} \lambda \, e^{- i \bm{p}_{\lambda}  \bm{\lambda}}
  \int d^{3} \rho \, e^{- i \bm{p}_{\rho}  \bm{\rho}}
       \Psi ( \lambda , \, \rho ) ,
\end{equation}
and its normalization is
\begin{equation}
  \int \frac{d^{3} p_{\lambda}}{( 2 \pi )^{3}}
  \int \frac{d^{3} p_{\rho}}{( 2 \pi )^{3}} 
  \left | \tilde{\Psi} ( p_{\lambda} , \, p_{\rho} ) \right | ^{2} = 1 .
\end{equation}

Finally, taking into account the antisymmetrization for the nucleons
in ${}^{3} \text{He}$, we express the full wave function of the
${}^{3} \text{He}$ nucleus as
\begin{align}
  \left | {}^{3} \text{He} ( \chi ) \right \rangle
  = \frac{1}{\sqrt{6}} \tilde{\Psi} ( p_{\lambda} , \, p_{\rho} )
  \big [ &
  | n ( p_{1} , \, \chi )
  p ( p_{2} , \, \chi_{\uparrow} )
  p ( p_{3} , \, \chi_{\downarrow} ) \rangle
  -
  | n ( p_{1} , \, \chi )
  p ( p_{3} , \, \chi_{\downarrow} )
  p ( p_{2} , \, \chi_{\uparrow} ) \rangle
  \notag
  \\
  & 
  - | p ( p_{2} , \, \chi_{\uparrow} )
  n ( p_{1} , \, \chi )
  p ( p_{3} , \, \chi_{\downarrow} ) \rangle
  + | p ( p_{3} , \, \chi_{\downarrow} )
  n ( p_{1} , \, \chi )
  p ( p_{2} , \, \chi_{\uparrow} ) \rangle
  \notag
  \\
  &
  + | p ( p_{2} , \, \chi_{\uparrow} )
  p ( p_{3} , \, \chi_{\downarrow} )
  n ( p_{1} , \, \chi ) \rangle
  - | p ( p_{3} , \, \chi_{\downarrow} )
  p ( p_{2} , \, \chi_{\uparrow} )
  n ( p_{1} , \, \chi ) \rangle
  \big ]
,
  \label{eqA:WFspin}
\end{align}
for the spinor of the $\HeT$, $\chi = \chi_{\uparrow} = ( 1 , \, 0
)^{t}$ or $\chi_{\downarrow} = ( 0 , \, 1 )^{t}$.  In this study the
spin direction of $\HeT$ is taken to be the same as that of the
neutron.

\section{$K^{-} p \to \bar{K}^{0} n$ and $K^{-} n \to K^{-} n$
  scattering amplitudes at $k_{\rm lab} = 1 \gev /c$}
\label{app:2}

In this Appendix we formulate the $K^{-} p \to \bar{K}^{0} n$ and
$K^{-} n \to K^{-} n$ scattering amplitudes at $k_{\rm lab} = 1 \gev
/c$, which is needed to emit the fast neutron in the final state of
the ${}^{3} \text{He} ( K^{-}, \, \Lambda p ) n$ reaction.  In this
study, we neglect the spin flip contribution and estimate the
scattering amplitude $T_{\bar{K} N \to \bar{K} N}$ of these reactions
at $k_{\rm lab} = 1 \gev / c$ from the differential cross section $d
\sigma _{\bar{K} N \to \bar{K} N} / d \Omega$ with the following
formula
\begin{equation}
  T_{\bar{K} N \to \bar{K} N} ( \cos \theta )
  = \frac{4 \pi w_{1}}{m_{N}}
  \sqrt{\frac{d \sigma _{\bar{K} N \to \bar{K} N}}{d \Omega}} ,
\end{equation}
where $\theta$ is the scattering angle for the kaon, $w_{1} \equiv
\sqrt{(p_{\bar{K}} + p_{N})^{2}}$, and $m_{N}$ is the nucleon mass.
When we theoretically calculate the differential cross section, we
always fix the initial kaon momentum as $k_{\rm lab} = 1 \gev /c$.

\begin{table}[t]
  \caption{Parameter sets for the $K^{-} p \to \bar{K}^{0} n$ and
    $K^{-} n \to K^{-} n$ scattering amplitudes at $k_{\rm lab} = 1
    \gev /c$.  All the parameters are given in units of mb/sr.}
  \label{tabA:amp_para}
  \centering
  \begin{tabular}{lcc}
    \hline
    \hline
    & $K^{-} p \to \bar{K}^{0} n$ &
    $K^{-} n \to K^{-} n$
    \\
    \hline
    $c_{0}$ & $0.57$ & $0.94$
    \\
    $c_{1}$ & $-0.05\phantom{-}$ & $0.51$
    \\
    $c_{2}$ & $0.65$ & $1.38$
    \\
    $c_{3}$ & $-0.42\phantom{-}$ & ---
    \\
    $c_{4}$ & $0.76$ & ---
    \\
    \hline
    \hline
  \end{tabular}
\end{table}

\begin{figure}[b]
  \centering
  \Psfig{7.7cm}{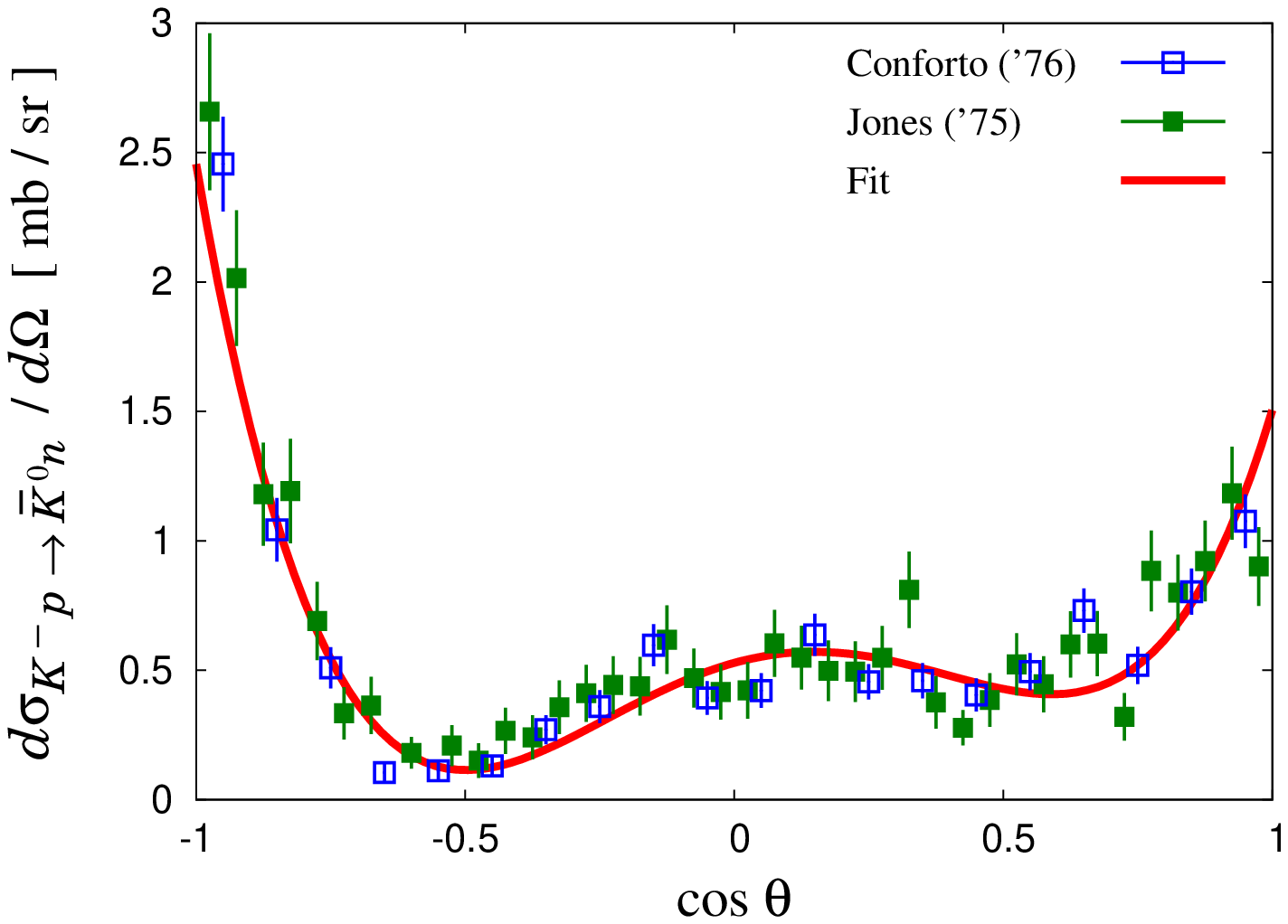}~\Psfig{7.7cm}{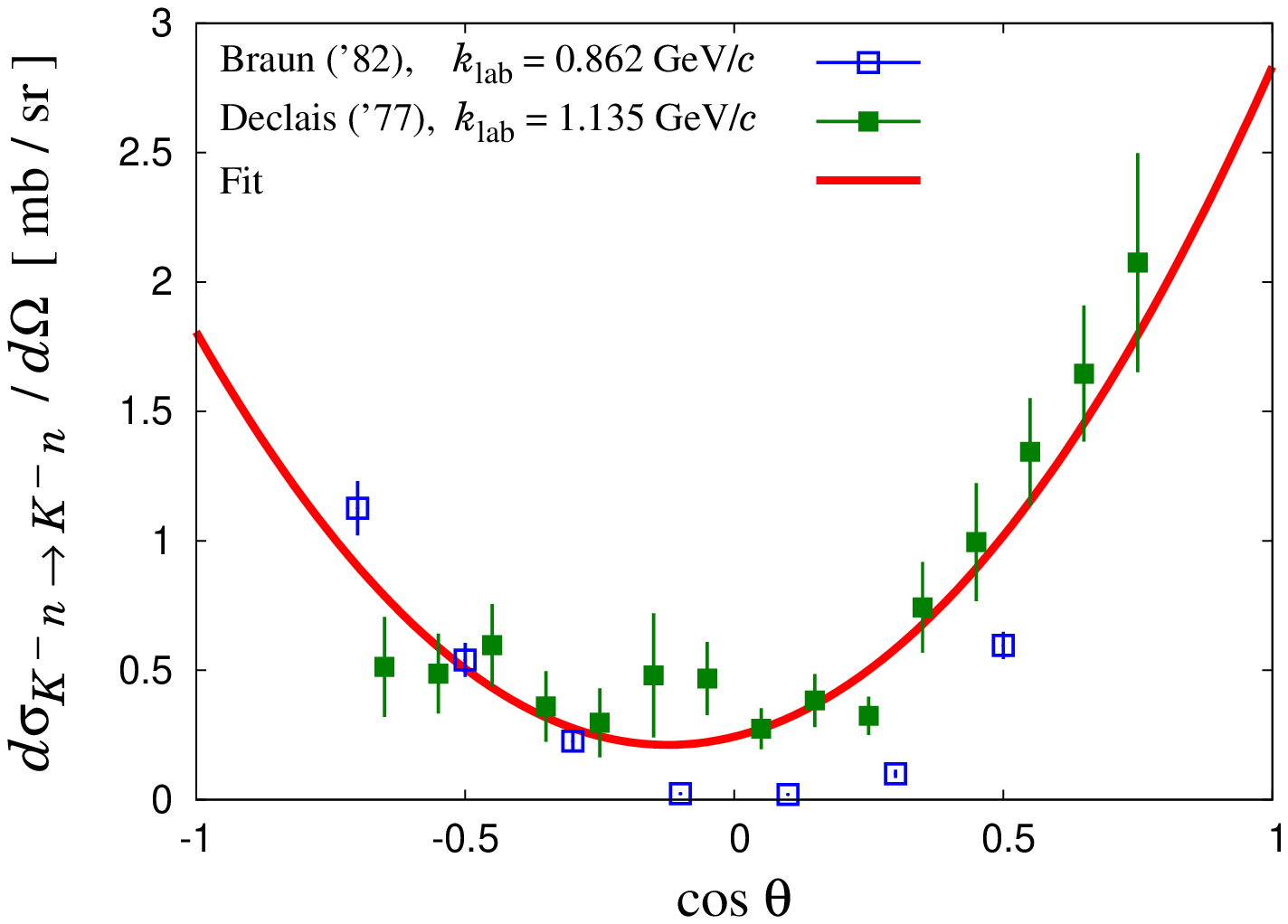}
  \caption{Differential cross sections of the $K^{-} p \to \bar{K}^{0}
    n$ (left) and $K^{-} n \to K^{-} n$ (right) reactions, where
    $\theta$ is the angle of the emerging kaon versus the original one.
    The experimental data are taken from Refs.~\cite{Jones:1974at,
      Conforto:1975nw} at $k_{\rm lab} = 1 \gev /c$ for a proton
    target, and from Ref.~\cite{Declais:1977kj} at $k_{\rm lab} =
    1.138 \gev /c$ and from~\cite{Braun:1982ih} at $k_{\rm lab} =
    0.862 \gev / c$ for a neutron target.}
  \label{figA:dsdO}
\end{figure}

We parametrize the differential cross section $d \sigma _{\bar{K} N
  \to \bar{K} N} / d \Omega$ by the Legendre polynomials $P_{l} ( x )$
as
\begin{equation}
  \frac{d \sigma _{\bar{K} N \to \bar{K} N}}{d \Omega}
  = \sum _{l} c_{l} P_{l} ( \cos \theta ) ,
\end{equation}
with constants $c_{l}$, which are fixed so as to reproduce the
experimental data.  For the proton target reaction, we have many
experimental data points for the differential cross section at $k_{\rm
  lab} = 1 \gev /c$~\cite{Jones:1974at, Conforto:1975nw}, so we take
the polynomials up to $l = 4$.  From the best fit we obtain the
parameters $c_{l}$ listed in Table~\ref{tabA:amp_para}.  For the
neutron target reaction, on the other hand, only the data at $k_{\rm
  lab} = 1.138 \gev /c$~\cite{Declais:1977kj} and at $k_{\rm lab} =
0.862 \gev / c$~\cite{Braun:1982ih} are available, so we take the
polynomials up to $l = 2$ and make a rough fit to the cross sections
at these momenta.  As a result, we obtain the parameters in
Table~\ref{tabA:amp_para}.  For both reactions, the fitted
differential cross sections are shown in Fig.~\ref{figA:dsdO} together
with the experimental data.

\section{$\bar{K} N \to \bar{K} N$ scattering amplitude around threshold}
\label{app:3}

In this Appendix we briefly introduce the $\bar{K} N \to \bar{K} N$
scattering amplitude around threshold, which appears in the secondary
scattering of the $\HeT ( K^{-} , \, \Lambda p ) n$ reaction and in
the multiple scattering of the $\bar{K} N N$ system.  For this
amplitude we employ the so-called chiral unitary
approach~\cite{Kaiser:1995eg, Oset:1997it, Oller:2000fj, Jido:2003cb},
modified to take into account the kaon absorption by two nucleons in
the $\bar{K} N N$ system in a simple way.  In this study we introduce
ten meson--baryon channels: $K^{-} p$, $\bar{K}^{0} n$, $\pi ^{0}
\Lambda$, $\pi ^{0} \Sigma ^{0}$, $\pi ^{+} \Sigma ^{-}$, $\pi ^{-}
\Sigma ^{+}$, $\eta \Lambda$, $\eta \Sigma ^{0}$, $K^{0} \Xi ^{0}$,
and $K^{+} \Xi ^{-}$.  Since we are interested in the amplitude around
the $\bar{K} N$ threshold, we only consider its $s$-wave part.

\begin{figure}[t]
  \centering
  \begin{tabular}{ccc}
    \PsfigII{0.25}{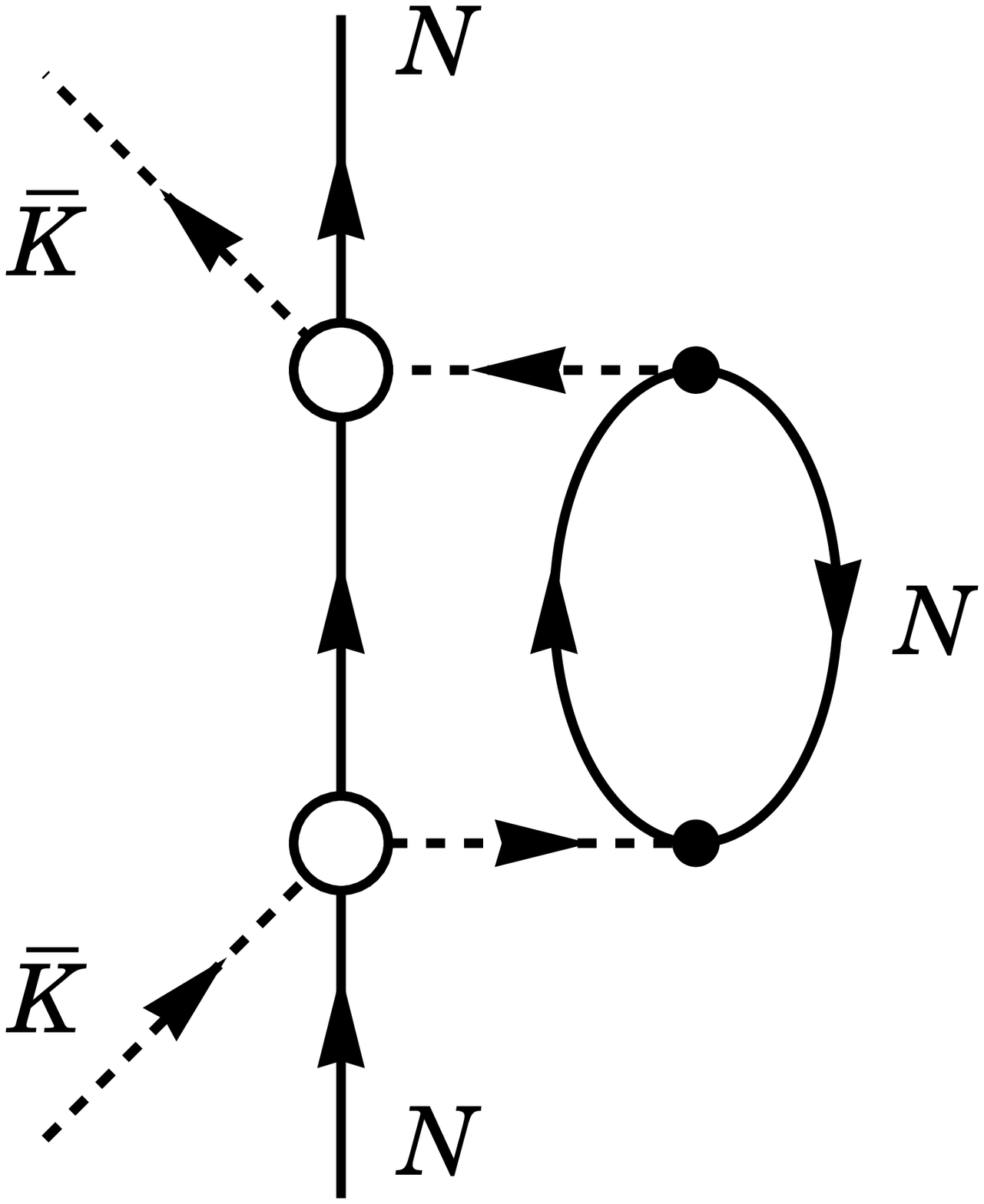} & & \Psfig{8.6cm}{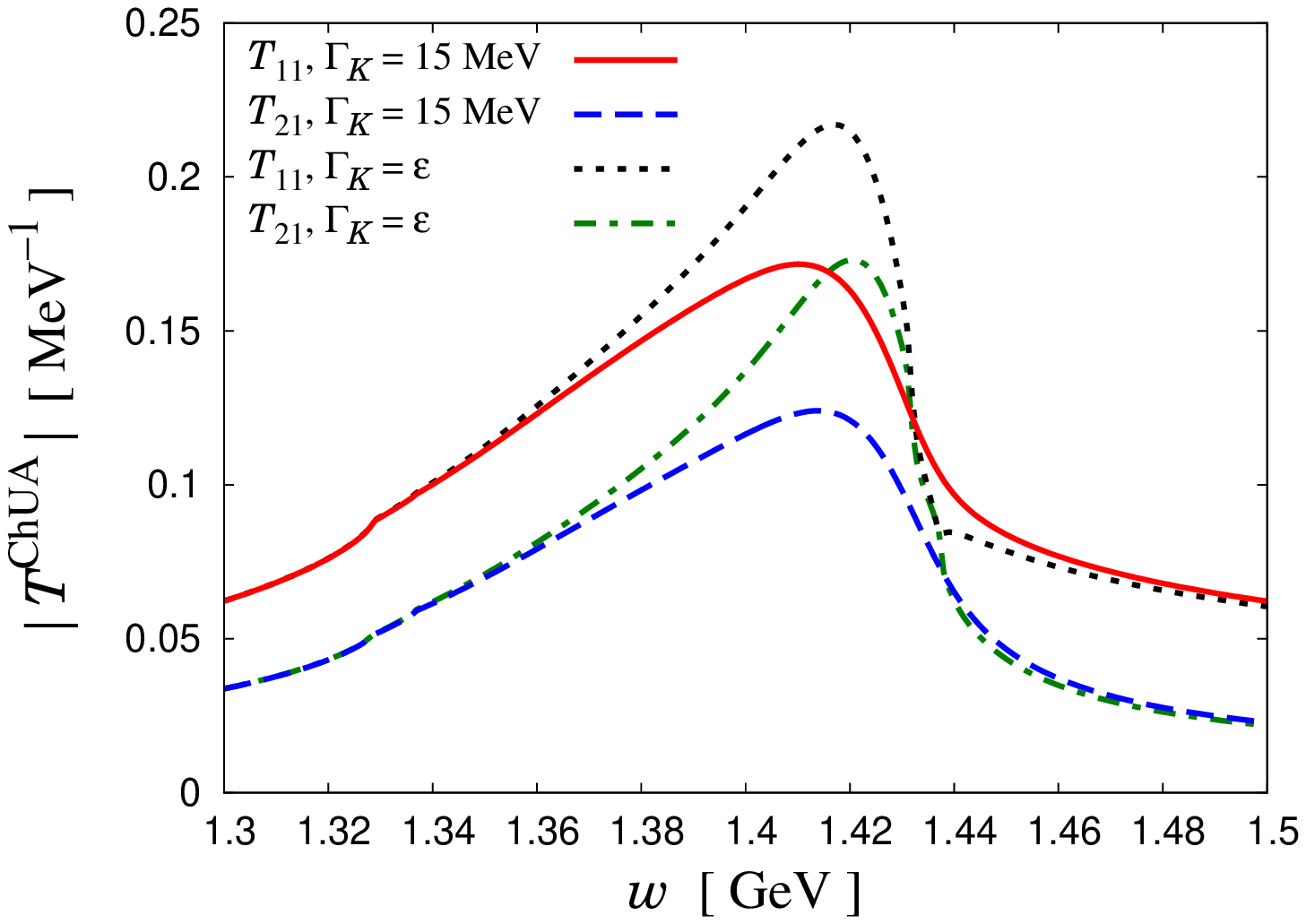} \\
    (a) & & (b) 
  \end{tabular}
  \caption{(a) Feynman diagram corresponding to the kaon absorption by
    two nucleons.  The unspecified solid and dashed lines represent
    baryons and mesons, respectively.  (b) Absolute values of the
    scattering amplitudes in chiral unitary approach for the $K^{-} p
    \to K^{-} p$ ($T_{11}$) and $\bar{K}^{0} n \to K^{-} p$ ($T_{21}$)
    reactions with the kaon absorption width $\Gamma _{K} = 15 \mev$.
    For comparison we also show the amplitudes with zero absorption
    width, $\Gamma _{K} = \epsilon$.}
  \label{fig:app3}
\end{figure}

In the chiral unitary approach, the $\bar{K} N \to \bar{K} N$
scattering amplitude $T_{i j}^{\rm ChUA}$, where $i$ and $j$ are
channel indices, is the solution of the coupled-channels
Lippmann-Schwinger equation in an algebraic form:
\begin{equation}
  T_{i j}^{\rm ChUA} ( w ) = V_{i j} ( w )
  + \sum _{k} V_{j k} ( w ) G_{k} ( w ) T_{k j}^{\rm ChUA} ( w )
  = \sum _{k} { [ 1 - V G ]^{-1}}_{i k} V_{k j} ,
\end{equation}
Here, $w$ is the center-of-mass energy, $V_{i j}$ is the interaction
kernel taken from chiral perturbation theory, and $G_{k}$ is the
meson-baryon loop function.  In this study the interaction kernel
$V_{i j}$ is fixed to be the leading order term of chiral perturbation
theory for the $s$-wave meson-baryon scattering, i.e., the
Weinberg-Tomozawa term, whose expression is given in
Ref.~\cite{Jido:2003cb}.  We note that in this scheme we can effectively take
into account the kaon absorption by two nucleons, whose diagram is shown in
Fig.~\ref{fig:app3}(a), in a simple way by just adding the imaginary
part of that diagram by means of an empirical width  $\Gamma _{K}$ in the kaon propagator. The loop function is then evaluated with the cutoff scheme as
\begin{equation}
  G_{i} ( w ) = \int _{q < q_{\rm max}} \frac{d^{3} q}{( 2 \pi )^{3}}
  \frac{M_{i}}{2 \omega _{i} ( \bm{q} ) E_{i} ( \bm{q} )}
  \frac{1}{w - \omega _{i} ( \bm{q} ) - E_{i} ( \bm{q} ) + i \Gamma _{i} } ,
\end{equation}
\begin{equation}
  \omega _{i} ( \bm{q} ) \equiv \sqrt{ \bm{q}^{2} + m_{i}^{2} } ,
  \quad
  E_{i} ( \bm{q} ) \equiv \sqrt{ \bm{q}^{2} + M_{i}^{2} } ,
  \quad
  \Gamma _{i} =
  \begin{cases}
    \Gamma _{K} & \text{for } i = K^{-} p , \, \bar{K}^{0} n , \\
    \epsilon & \text{for other channels} ,
  \end{cases}
\end{equation}
where $q_{\rm max}$ is the cutoff, $M_{i}$ and $m_{i}$ are the baryon and
meson masses in channel $i$, respectively, $\Gamma _{K}$ is the kaon
absorption width by two nucleons, and $\epsilon$ is an infinitesimal
positive value.  

In this study we take the cutoff as $q_{\rm max} = 630
\mev$~\cite{Oset:1997it}, and the kaon absorption width is fixed to be
$\Gamma _{K} = 15 \mev$ so as to reproduce the kaon absorption width
of the $\bar{K} N N$ bound state in the fixed center
approximation~\cite{Bayar:2012hn} (see Appendix~\ref{app:4}).  With
these values, we can calculate the $\bar{K} N \to \bar{K} N$
scattering amplitude as plotted in Fig.~\ref{fig:app3}(b).  As one can
see from the figure, the introduction of the absorption width $\Gamma
_{K} = 15 \mev$ shifts the peak position for the $\Lambda (1405)$
resonance $\sim 1.4 \gev$ only slightly, but the width of the $\Lambda
(1405)$ peak grows to $\sim 50 \mev$.  Actually, for this
amplitude with $\Gamma _{K} = 15 \mev$, we find a resonance pole
corresponding to the peak in Fig.~\ref{fig:app3}(b) at $w = 1428 - 26
i \mev$ in the complex energy plane, which was $1427 - 19 i \mev$ for
$\Gamma _{K} = \epsilon$.

\section{Kaon multiple scattering amplitude in the fixed center approximation}
\label{app:4}

In this Appendix we formulate the kaon multiple scattering amplitude
for the $\bar{K} N N$ three-body state.  In this study we employ the
so-called fixed center approximation to the Faddeev
equation~\cite{Bayar:2011qj, Bayar:2012hn}.  For the sake of
simplicity in the calculation, isospin symmetric masses are employed
in the calculation of the $\bar{K} N N$ multiple scattering amplitude.

Since we are interested in the $K^{-} p p$ states and others coupling
to this, we consider six channels for the $\bar{K} N N$ three-body
state: $K^{-} p p$, $\bar{K}^{0} n p$, $\bar{K}^{0} p n$, $p p K^{-}$,
$n p \bar{K}^{0}$, and $p n \bar{K}^{0}$, which are labeled by indices
$i$ and $j$ in the above order.  We note that we distinguish the
ordering of the kaon and two nucleons so as to specify the nucleon
with which the kaon interacts first and last.  For instance, in the
channel $K^{-} p n$ ($p n K^{-}$) in the initial state, $K^{-}$
interacts first with the proton on the left (neutron on the right).
With this, we can divide the multiple scattering processes into an
even- and odd-number of scatterings, which makes the formulation of
the multiple scattering easier.  

We note that pion exchange between two nucleons is neglected in the
diagrams that build the $T^{\rm FCA}$ amplitude, see
Fig.~\ref{fig:FCA}.  In fact, these pion exchange contributions may
appear either in the last of the multiple scatterings or may
correspond to exchanges in intermediate states. In the first case, the
pion exchange is accompanied by the $\bar{K} N \to \pi \Lambda$
amplitude and by the absorption of the pion via the $\pi N N$ vertex.
Since the amplitude $\bar{K} N \to \pi \Lambda$ has isospin $I = 1$,
the contribution from the first case should be small.  In the second
case, the pion exchange is accompanied by the $\pi N \to \pi N$
amplitude, but this contribution in the intermediate states has been
found to be small in Ref.~\cite{Kamalov:2000iy}.

\begin{figure}[!t]
  \centering
  \PsfigII{0.225}{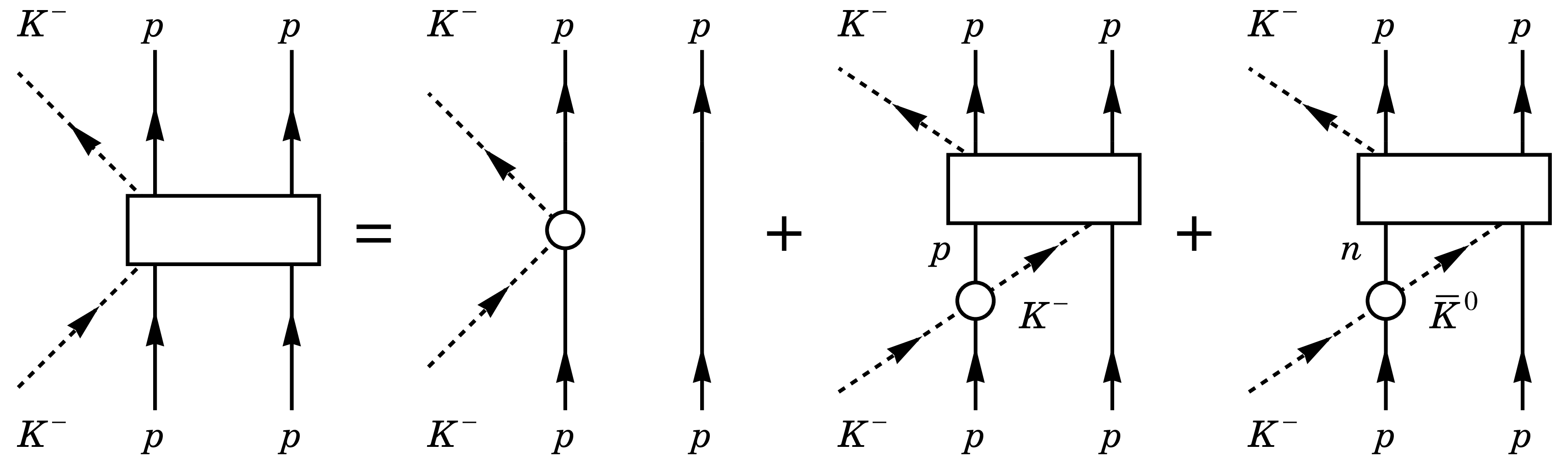} 
  \caption{Feynman diagrams for the kaon multiple scattering of the
    process $K^{-} p p \to K^{-} p p$.  The open rectangles indicate
    the $\bar{K} N N$ multiple scattering amplitude and the circles
    represent the $\bar{K} N \to \bar{K} N$ one.}
  \label{figA:Kpp}
\end{figure}

Here, in order to grasp the structure of the multiple scattering in
the fixed center approximation, we first consider the $K^{-} p p \to
K^{-} p p$ amplitude $T_{11}^{\rm FCA}$, which is diagrammatically
shown in Fig.~\ref{figA:Kpp}.  This can be expressed in the following
equation:
\begin{equation}
  T_{11}^{\rm FCA} = t_{1} 
  + t_{1} G_{0} T_{41}^{\rm FCA}
  + t_{2} G_{0} T_{51}^{\rm FCA} ,
\end{equation}
where $t$ is the $\bar{K} N \to \bar{K} N$ scattering amplitude and
$G_{0}$ is the kaon propagator, both of which are functions only of
the total energy of the $\bar{K} N N$ system, i.e., the $\Lambda p$
invariant mass.  The $\bar{K} N \to \bar{K} N$ scattering amplitude
is fixed in the chiral unitary approach
\begin{equation}
  t_{1} ( M_{\Lambda p} ) = T_{K^{-} p \to K^{-} p}^{\rm ChUA} ( w_{\rm FCA} ) ,
  \quad 
  t_{2} ( M_{\Lambda p} ) = T_{K^{-} p \to \bar{K}^{0} n}^{\rm ChUA} ( w_{\rm FCA} ) 
  = T_{\bar{K}^{0} n \to K^{-} p}^{\rm ChUA} ( w_{\rm FCA} ) ,
  \label{eqA:t1t2}
\end{equation}
where the argument $w_{\rm FCA}$ is~\cite{Bayar:2011qj}:
\begin{equation}
  w_{\rm FCA} ( M_{\Lambda p} ) \equiv
  \sqrt{\frac{M_{\Lambda p}^{2} + m_{K}^{2} - 2 m_{N}^{2}}{2}} .
\end{equation}
The kaon propagator $G_{0}$ is evaluated as
\begin{equation}
  G_{0} ( M_{\Lambda p} ) \equiv \int \frac{d^{3} q}{( 2 \pi )^{3}}
  \frac{F_{N N} ( \bm{q} )}{q_{\rm FCA}^{0} ( M_{\Lambda p} )^{2} 
    - \omega _{K} ( \bm{q} )^{2} + i m_{K} \Gamma _{K}} ,
\end{equation}
\begin{equation}
  q_{\rm FCA}^{0} ( M_{\Lambda p} ) \equiv
  \frac{M_{\Lambda p}^{2} + m_{K}^{2} - ( 2 m_{N} )^{2}}{2 M_{\Lambda p}} ,
\end{equation}
with the kaon energy $\omega _{K} ( \bm{q} ) \equiv \sqrt{\bm{q}^{2} +
  m_{K}^{2}}$ and the form factor for the $N N$ system:
\begin{equation}
  F_{N N} ( \bm{q} ) \equiv \int d^{3} r \, e^{i \bm{q}  \bm{r}}
  | \varphi ( r ) |^{2} .
\end{equation}
Here $\varphi ( r )$ is the wave function for the $N N$ system, for
which we take the $\rho$ mode of the $\HeT$ wave function as (see
Appendix~\ref{app:1}):
\begin{equation}
  F_{N N} ( \bm{q} ) = \int d^{3} \rho \, e^{i \bm{q}  \bm{\rho}}
  \left ( \frac{m_{\rho} \omega _{\rho}}{\pi} \right )^{3/2}
  e^{- m_{\rho} \omega _{\rho} \rho ^{2}} 
  = e^{- \bm{q}^{2} / ( 4 m_{\rho} \omega _{\rho} )} ,
\end{equation}
with $m_{\rho} = m_{N}/2$ and $\omega _{\rho}$ in
Eq.~\eqref{eqA:omega}.\footnote{For the $\HeT$ wave function, we use
  isospin symmetric nucleon mass $m_{N}$: $m_{N} = (m_{p} + m_{n}) /
  2$.}

In a similar manner we can express the multiple scattering amplitude
$T_{i j}^{\rm FCA}$ with the tree-diagram contributions, containing only $\bar{K} N$ two-body amplitudes, and tree times
further multiple scattering amplitudes.  As a result, the multiple
scattering amplitude $T_{i j}^{\rm FCA}$ is found to be a solution of
the scattering equation:
\begin{equation}
  T_{i j}^{\rm FCA} = V_{i j}^{\rm FCA}
  + \sum _{k = 1}^{6} \tilde{V}_{i k}^{\rm FCA} G_{0} T_{k j}^{\rm FCA}
  = \sum _{k = 1}^{6} {\left [ 1 - \tilde{V}^{\rm FCA} G_{0} \right ]^{-1}}_{i k}
  V_{k j}^{\rm FCA} 
\end{equation}
with
\begin{equation}
  V^{\rm FCA} =
  \left (
  \begin{array}{@{\,}cccccc@{\,}}
    t_{1} & t_{2} & 0 & 0 & 0 & 0 \\
    t_{2} & t_{3} & 0 & 0 & 0 & 0 \\
    0 & 0 & t_{4} & 0 & 0 & 0 \\
    0 & 0 & 0 & t_{1} & 0 & t_{2} \\
    0 & 0 & 0 & 0 & t_{4} & 0 \\
    0 & 0 & 0 & t_{2} & 0 & t_{3} \\
  \end{array}
  \right ) ,
  \quad
  \tilde{V}^{\rm FCA} =
  \left (
  \begin{array}{@{\,}cccccc@{\,}}
    0 & 0 & 0 & t_{1} & t_{2} & 0 \\
    0 & 0 & 0 & t_{2} & t_{3} & 0 \\
    0 & 0 & 0 & 0 & 0 & t_{4} \\
    t_{1} & 0 & t_{2} & 0 & 0 & 0 \\
    0 & t_{4} & 0 & 0 & 0 & 0 \\
    t_{2} & 0 & t_{3} & 0 & 0 & 0 \\
  \end{array}
  \right ) .
\end{equation}
Here $t_{3}$ and $t_{4}$ are
\begin{equation}
  t_{3} ( M_{\Lambda p} ) = T_{\bar{K}^{0} n \to \bar{K}^{0} n}^{\rm ChUA} ( w_{\rm FCA} ) ,
  \quad 
  t_{4} ( M_{\Lambda p} ) = T_{\bar{K}^{0} p \to \bar{K}^{0} p}^{\rm ChUA} ( w_{\rm FCA} ) ,
\end{equation}
and $t_{1}$ and $t_{2}$ have been defined in Eq.~\eqref{eqA:t1t2}.  We
note that the multiple scattering amplitude $T^{\rm FCA}$ depends only
on the invariant mass $M_{\Lambda p}$.

With this formulation, we can plot the multiple scattering amplitude
as in Fig.~\ref{fig:T_FCA}.  The kaon absorption width $\Gamma _{K}$
in $G_0$ is fixed as $\Gamma _{K} = 15 \mev$, so that the amplitude
$T_{11}^{\rm FCA} + T_{41}^{\rm FCA}$ reproduces the width of the
$\bar{K} N N$ bound-state signal in the fixed center
approximation~\cite{Bayar:2012hn}.  For this amplitude in the fixed
center approximation, we find a pole at $M_{\Lambda p} = 2354 - 36 i
\mev$ in the complex plane of the invariant mass $M_{\Lambda p}$,
corresponding to the peak in Fig.~\ref{fig:FCA}.

\end{document}